\documentclass[iop]{emulateapj}
\usepackage{apjfonts}
\pdfoutput=1

\newcommand{\ha}{\rm{H}\alpha}
\newcommand{\hb}{\rm{H}\beta}

\newcommand{\mzs}{{$M_*$--$Z$--SFR}}
\newcommand{\mz}{{$M_*$--$Z$}}

\shorttitle{MASS-METALLICITY-SFR RELATION}
\shortauthors{SALIM ET AL.}

\begin{document}

\title{On the Mass--Metallicity--Star Formation Rate
  Relation for Galaxies at $z\sim2$}

\author{Samir Salim\altaffilmark{1}, Janice C.\ Lee\altaffilmark{2,3},
  Romeel Dav\'e\altaffilmark{4}, Mark Dickinson\altaffilmark{5}
} 
\altaffiltext{1}{Department of Astronomy, Indiana University,
  Bloomington, IN 47404, USA, salims@indiana.edu} 
\altaffiltext{2}{Space Telescope Science Institute, Baltimore, MD 21218, USA} 
\altaffiltext{3}{Visiting Astronomer, Spitzer Science Center, Caltech,
  Pasadena, CA 91125, USA.}
\altaffiltext{4}{University of the Western Cape, Bellville, Cape Town
  7535, South Africa}
\altaffiltext{5}{National Optical Astronomy Observatory, Tucson, AZ
  85719}


\begin{abstract}
  Recent studies have shown that the local mass--metallicity (\mz)
  relation depends on the specific star formation rate (SSFR). Whether
  such a dependence exists at higher redshifts, and whether the
  resulting \mzs\ relation is redshift invariant, is debated. We
  re-examine these issues by applying the non-parametric techniques of
  Salim et al.\ (2014) to $\sim130$ $z\sim2.3$ galaxies with N2 and O3
  measurements from KBSS (Steidel et al.\ 2014). We find that the KBSS
  \mz\ relation depends on SSFR at intermediate masses, where such
  dependence exists locally. KBSS and SDSS galaxies of the same mass
  and SSFR (``local analogs'') are similarly offset in the BPT diagram
  relative to the bulk of local star-forming galaxies, and thus we
  posit that metallicities can be compared self-consistently at
  different redshifts as long as the masses and SSFRs of the galaxies
  are similar.  We find that the \mzs\ relation of $z\sim2$ galaxies
  is consistent with the local one at $\log M_*<10$, but is offset up
  to $-0.25$ dex at higher masses, so it is altogether not redshift
  invariant. This high-mass offset could arise from a bias that
  high-redshift spectroscopic surveys have against high-metallicity
  galaxies, but additional evidence disfavors this possibility. We
  identify three causes for the reported discrepancy between N2 and
  O3N2 metallicities at $z\sim2$: (1) a smaller offset that is also
  present for SDSS galaxies, which we remove with new N2 calibration,
  (2) a genuine offset due to differing ISM condition, which is also
  present in local analogs, (3) an additional offset due to
  unrecognized AGN contamination.
\end{abstract}


\section{Introduction}

The relationship between the stellar masses, gas-phase metallicities,
and star formation rates of galaxies (\mzs) has received increasing
attention over the past $\sim$5 years, both because of its conceptual
simplicity and its potential to provide deep insight into the
processes that regulate galactic star formation (SF) and drive galaxy
evolution over cosmic time.  While the local ($z\lesssim0.3$)
luminosity-metallicity and mass-metallicity (\mz) relations have been
studied for decades, beginning with \citet{lequeux79}, only more
recently has the star formation rate (SFR) been proposed as a second
parameter in the local mass-metallicity relation \citep{ellison08}.
The primary correlation shows metallicity increasing with stellar
mass\footnote{Stellar masses are expressed in units of solar mass
  ($M_{\odot}$).} until a plateau is reached at log $M_*\sim$10.5
\citep{tremonti04}, while the general sense of the secondary
dependence with SFR is that at fixed mass, galaxies with higher SFRs
tend to be more metal poor
\citep{ellison08,mannucci10,lara-lopez10,hunt12}.

Conflicting results on the characterization of the local \mzs\
relation have ignited debate over whether the reported secondary
dependence on SFR could be spurious, and due to: sample selection
effects \citep{mithi}, correlated errors in the measurements of SFR
and metallicity \citep{lilly13}, systematic errors in metallicities
\citep{yates12}, or biases introduced by SDSS fiber spectroscopy
\citep{sanchez13}. In \citet{mithi}, the limitations of previous
parameterizations used to characterize the local relation and the need
for non-parametric techniques to enable comparative analysis of different
datasets were also highlighted.  

Thus, to gain insight into the origins of the conflicting results, in
\citet{s14} we devised a non-parametric analysis framework based on
the SFR offset from the local star-forming (``main'') sequence at a
given $M_*$, and undertook a comprehensive re-analysis of the local
\mzs\ relation using a Sloan Digital Sky Survey (SDSS) dataset
together with {\it GALEX} ultraviolet and {\it WISE} infrared
photometry.  Studying the \mzs\ relation in terms of SFRs or specific
SFRs (SSFRs) relative to typical values on the ``main sequence'' is
more physically motivated than using absolute SFRs, which, to first
order, scale with $M_*$ \citep{b04}. Although we concluded that the
dependence on SSFR is not spurious (after investigating multiple SFR
and metallicity indicators), the analysis exposed important features
of the relationship. In particular, \citet{s14} showed that adding the
SFR as a second parameter does not greatly decrease the scatter in the
\mz\ relation when the metallicities of individual galaxies, rather
than the median-binned values are considered (i.e., the \mzs\ relation
is not tight). We confirmed that the overall SFR dependence is weaker,
or absent, at higher masses. However, at a given mass the dependence
on SFR is much stronger for intensely star-forming galaxies above the
``main sequence'' (galaxies with high SSFR for their mass). We noted
that simple parameterizations of the local \mzs\ relation (a plane, or
the projection of least scatter in $Z$) do not capture this behavior
of galaxies with high relative SSFR because the parametrizations are
dominated by the bulk of ``normal" galaxies along the core of the main
sequence.

Recognizing the limitations of parameterizations of the local \mzs\ is
particularly important in the context of testing its redshift
invariance.  The concept of invariant \mzs\ relation was put forward
by \citet{mannucci10}, who refer to a it as the ``fundamental
metallicity relation'' (FMR).  They showed that selected galaxy
samples up to $z\sim2.2$ lie along the projection of the local \mzs\
that minimizes the scatter in metallicity, i.e., that they are
consistent with a non-evolving \mzs\ relation. \mz\ relations at $z>0$
then represent the slices of the invariant (i.e., fundamental) \mzs\
relation at (S)SFRs higher than those found in local
galaxies. However, a number of more recent studies have concluded that
$z\sim2$ samples do not lie on the local \mzs\ relation, implying an
evolving relation. Most of these studies were based on
\citet{mannucci10} parametrization of the local relation (e.g.,
\citealt{cullen14,zahid14b,maier14}), except for \citet{sanders15},
who performed a direct comparison with the local samples. The result
of \citet{sanders15} highlights other possible causes for the
discrepant results, e.g., an evolution in metallicity calibrations
(e.g., \citealt{kewley13,steidel14}). Sample selection effects
\citep{juneau14}, may also play a role. Establishing the existence of
SFR dependence of \mz\ relation at higher redshifts would be important
even if the resulting high-redshift \mzs\ relation did not coincide
with the one followed by local galaxies. The evidence that SFR is a
second parameter at $0.7<z<2.3$ is likewise inconclusive
\citep{cresci12,zahid14b,wuyts14,steidel14,maier14,mithi,sanders15}.
 
Armed with a more detailed, non-parametric characterization of the
local \mzs\ relationship we can examine these points of contention in
recent work involving galaxies at $z\sim2$.  Here, we will apply our
analysis framework to re-examine whether a secondary dependence of the
\mz\ relation on SFR also exists for higher redshift samples, and more
generally, whether the local \mzs\ relation is invariant with redshift
and describes star-forming galaxies at all stages of their evolution
over cosmic time. We focus on the latest and most comprehensive
$z\sim2$ spectroscopic datasets \citep{steidel14,sanders15}. We
present evidence that this approach is able to circumvent the thorny
issues of inconsistent or evolving metallicity calibrations. We
highlight the possible role of observational selection effects
\citep{juneau14} and/or the possibility that some high-mass
high-redshift galaxies contain unrecognized contribution from AGN line
emission.

\section{Data and samples} \label{sec:data}

Our analysis is primarily based on the dataset published in
\citet{steidel14} from the Keck Baryonic Structure Survey (KBSS), a
near-IR spectroscopic survey performed with the MOSFIRE multi-object
slit spectrograph on the Keck I telescope. KBSS has obtained $H$ and
$K$-band spectra of galaxies at $1.95<z<2.65$. The majority of targets
had previously determined redshifts from optical spectroscopy and had
been originally selected using a variety of techniques based on
rest-frame UV colors \citep{adelberger04,steidel04}. The
\citet{steidel14} analysis includes individual galaxies where the S/N
ratios of [OIII]5007 and $\ha$ line fluxes exceed 5, and those of
[NII]6584 and $\hb$ exceed 2. For an average exposure time of 11000 s,
these cuts corresponds to $\ha$ and [OIII] $5\sigma$ limits of $\sim
10^{41.7}$ erg s$^{-1}$ (obtained from scaling MOSDEF sensitivity
given by \citealt{coil15}). The published version of \citet{steidel14}
presents 161 galaxies with both N2 = log[$F$([NII]6584)/$F(\ha)$] and
O3 = log[$F$([OIII]5007)/$F(\hb)]$ line ratios and an additional 31
galaxies with only the N2 measurements (i.e., currently lacking
$H$-band observations). The total of 192 galaxies excludes seven
objects identified as AGN based on broad emission lines or the
presence of higher ionization species. Stellar masses were derived
from SED fitting, and SFRs from $\ha$ fluxes, corrected for slit
losses and for extinction based on the continuum dust attenuation
estimate from the SED fits.  We note that the data used in our
analysis, however, are the subset of the \citet{steidel14} sample,
which was presented in the preprint version of their paper. We use the
smaller, preprint sample because the published version of the paper
omitted SFRs from the tables. Thus, the dataset used here contains 108
galaxies with both N2 and O3, and an additional 18 with N2, i.e., 2/3
of the published sample. Line ratios and stellar masses are taken from
the published tables, and SFRs from the preprint. Differences between
preprint and published line ratios are not significant (0.07 dex
scatter, with no systematic offsets), so we assume that the SFRs have
not changed significantly either, as also evidenced by a similar
appearance of figures that involve SFRs in the preprint and the
published paper.

We augment the analysis with measurements from the MOSFIRE Deep
Evolution Field (MOSDEF) survey early observations
\citep{kriek14,coil15,shapley15}. MOSDEF targets $H$-band selected
galaxies at $2.1<z<2.6$ within the CANDELS survey areas. Targeting
prioritization criteria include the availability of a spectroscopic
redshift from previous work (40\% of the sample), brightness, and
photometric redshift in the target range. O3N2 ($=$ O3$-$N2) and
N2-based metallicities were obtained for 53 galaxies for which all
four lines have S/N ratio $>3$, corresponding to $\ha$ and [OIII]
$3\sigma$ limits of $\sim 10^{41.5}$ erg s$^{-1}$
\citep{coil15}\footnote{While KBSS is on average deeper than MOSDEF,
  the former requires higher detection threshold in [OIII]
  line.}. AGNs were removed from this sample based on X-ray and IR
indicators, or if N2$>-0.3$ \citep{coil15}. Stellar masses come from
SED fitting, and SFRs from $\ha$ fluxes corrected for extinction using
the Balmer decrement.  Data tables with measurements for individual
MOSDEF galaxies are not yet published. However, we used the published
figures in \citet{coil15} and \citet{sanders15} to compare \mz,
SSFR--$M_*$ and O3--N2 relations from MOSDEF with those from SDSS
and KBSS.

\begin{figure*}
\epsscale{1.2} \plotone{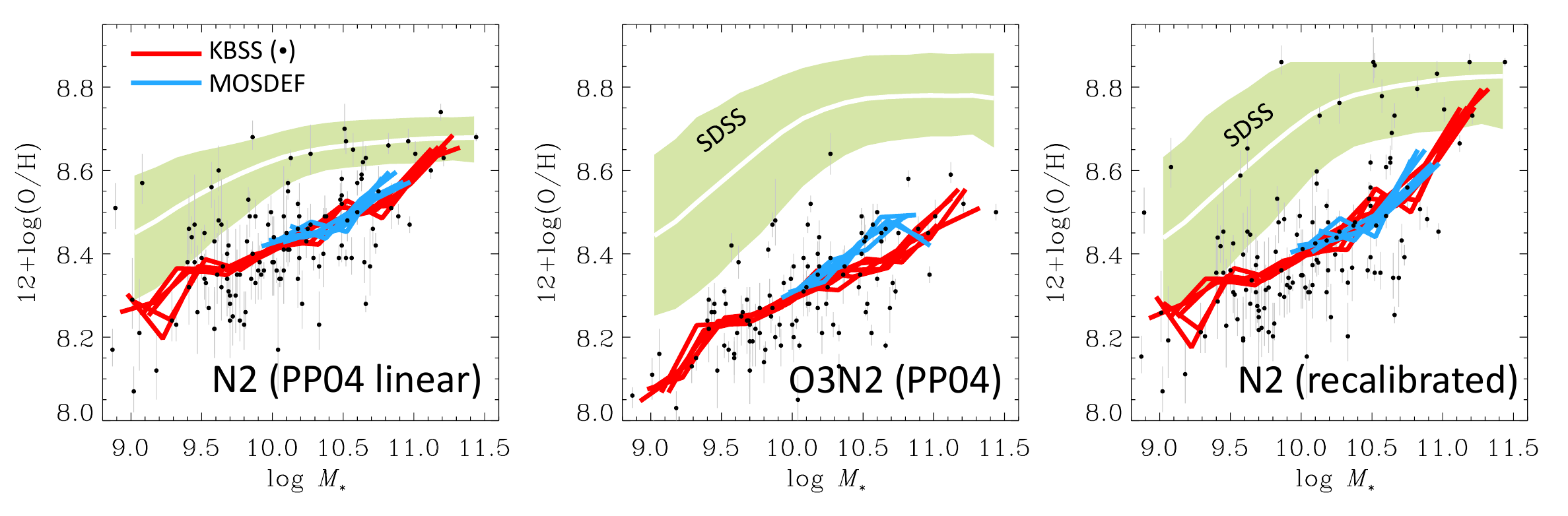}
\caption{The impact of the choice of metallicity indicator on inferred
  evolution in the \mz\ relation. Local ($z\sim0$) sample comes from
  SDSS (green band is the 95 percentile range at a given mass; white
  line shows an average trend), while $z\sim2$ data are from KBSS
  (black points; binned averages shown as red lines) and MOSDEF
  (binned averages shown as blue lines). MOSDEF averages are based on
  galaxies with $M_*\gtrsim10$, for which their sample is not affected
  by incompleteness. Different average lines are obtained by varying
  the starting position for binning in fifths of the 0.25 dex bin
  width. Left (middle) panel is based on N2 (O3N2) line ratios,
  converted into metallicity using \citet{pp04} linear
  calibrations. Right panel shows N2 after the recalibration is
  performed in order to bring N2 and O3N2 metallicities into a better
  mutual agreement for the SDSS sample. O3N2 shows a higher degree of
  ``evolution'' than N2, especially at higher masses. MOSDEF and KBSS
  MZRs agree between each other for N2, but depart to some extent for
  the more massive galaxies in the case of O3N2. \label{fig:mzr1}}
\end{figure*}

Comparison with the relations followed by local galaxies is based on
the SDSS DR7 spectroscopic sample \citep{sdsssp, abazajian09}
processed by the MPA/JHU group. Sample selection follows that of
\citet{mannucci10}, except that we extend the redshift range
($0.005<z<0.3$) as long as the mass included in the spectroscopic
fiber is $>10\%$, following what we have done in \cite{s14}.  Dropping
the redshift limit from 0.07 to 0.005 removes low-SFR incompleteness
for star-forming galaxies with $9<\log M_*<10$ \citep{s14}. Typical
mass-covering fraction of SDSS fiber spectroscopy is 30\%, with 95
percentile range between 17\% and 50\%. Inclusion of lower-redshift
galaxies, which have lower mass-covering fraction, does not affect any
of the results.  It is also important to note that the selection is
based solely on S/N ratio in $\ha$ being above 25. The limit on only
the $\ha$ line ensures that the sample is not biased in metallicity
\citep{s14,mannucci10}, while it is high enough that other required
nebular lines will be well measured. Lowering the limit to values
below 25 increases the number of galaxies with lower relative
SSFR. However, these galaxies entirely follow the trends established
by galaxies with higher S/N ratio, just with less precise metallicity
measurements.

For initial analysis, we exclude SDSS galaxies that lie above the
\citet{k03c} AGN demarcation line. We use total (integrated) SFRs and
stellar masses from the MPA/JHU catalog, which are determined
following \citet{b04} and \citet{s07}, respectively, with additional
details given in online
documentation.\footnote{\url{http://www.mpa-garching.mpg.de/SDSS/DR7}.}
Specifically, \citet{b04} SFRs are based on a hybrid combination of an
emission-line based SFR within the spectral fiber and a photometric
estimate outside of the fiber, and hence should capture the total
activity of the galaxies.

A Chabrier IMF and standard cosmology ($H_0=70$ km s$^{-1}$
Mpc$^{-1}$, $\Omega_m=0.3$, $\Omega_{\Lambda}=0.7$) is assumed
throughout the paper.

\section{Method}

As in \citet{s14}, we examine the metallicity as a function of the 
relative SSFR at fixed stellar mass. 
We define the relative SSFR as the offset from the {\it local}
(SDSS) star-forming (``main'') sequence:

\begin{equation}
\Delta \log {\rm SSFR} = \log {\rm SSFR} - \langle\log {\rm SSFR}\rangle_{M_*},
\end{equation}

\noindent where $\langle\log {\rm SSFR}\rangle_{M_*}$ is the median
$\log {\rm SSFR}$ of galaxies with $M_*$. Alternatively (not used
here), $\langle\log {\rm SSFR}\rangle_{M_*}$ can be a value obtained
from fitting the local ``main'' sequence with some mean relation (e.g.,

\begin{equation}
\langle\log {\rm SSFR} \rangle_{M_*}= -0.35(\log M_*-10)-9.83
\label{eqn:ssfr_mass}
\end{equation}

\noindent from \citealt{s07}). We define relative SSFRs in reference
to local main sequence (as opposed to high-redshift one) because the
local SSFR--$M_*$ relation is more robustly known. In principle,
relative SSFRs can be defined with respect to high-redshift sequence
as well, and this change in zero point would not impact the analysis.

Using this simple analysis framework, one can determine if an SFR
dependence is present without assuming a parametrization of the \mzs\
relation.  Such non-parametric techniques are essential because they
allow the dependence on the mass and SFR to be simultaneously
explored, in contrast to a plane parameterization of
\citet{lara-lopez10}, which forces a fixed dependence on both the mass
and the SFR, or the projection of least scatter in $Z$ of
\citet{mannucci10}, which forces a fixed SFR dependence at a given
mass. Moreover, our methodology allows one to test whether
high-redshift samples follow the same \mzs\ relation as the local
galaxies without extrapolation of the assumed parametrization into
regions which may not be well populated by local galaxies, and
therefore do not carry much weight in the parameterization, but are
occupied by high-redshift galaxies.

We will also apply a related non-parametric method for an even more
direct test of \mzs\ invariance. The method consists in comparing the
metallicities of $z\sim2.3$ galaxies with the metallicities of their
{\it local ``analogs''}. A local ``analog'' is defined as an SDSS
galaxy with $M_*$ and SSFR most similar to a given high-redshift
galaxy, i.e., a galaxy for which the metric:

\begin{equation}
D^2 = (\Delta \log {\rm SSFR})^2 +  (\Delta\log M_*)^2
\label{eqn:metric}
\end{equation}

\noindent is minimized. Large volume of SDSS allows finding very close
matches in SSFR and $M_*$ (small $D$). We use the word ``analog'' with
caution, because we are not implying that such local galaxies undergo
the same physical processes as the high-redshift galaxies, but only
that in the context of invariant \mzs\ relation, the metallicities of
high-redshift galaxies and the local ``analogs'' should be the same.

\begin{figure*}
\epsscale{1.2} \plotone{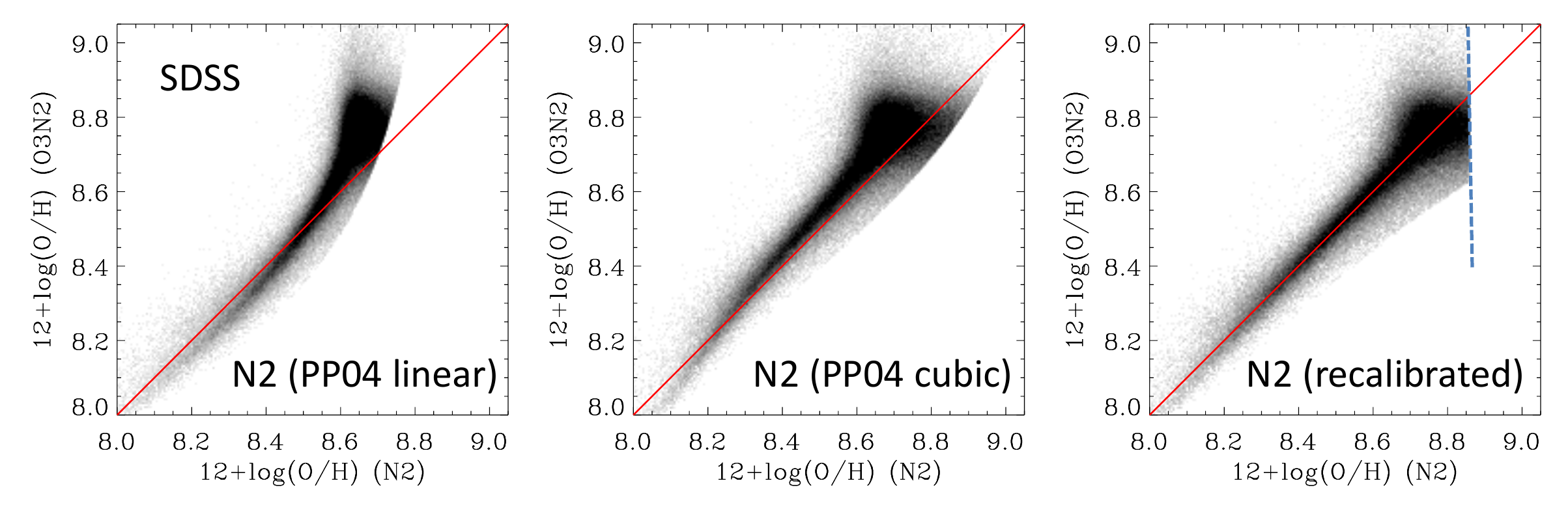}
\caption{Comparison of metallicities of SDSS galaxies derived based on
  O3N2 and N2 line ratios. O3N2 metallicity in all panels is derived
  from \citet{pp04} calibration, while N2 metallicity is based on
  linear and cubic \citet{pp04} relations (left and middle). One
  expects \citet{pp04} calibrations, being local, to yield consistent
  metallicities for SDSS galaxies, but this is not entirely the
  case. We therefore recalibrate N2 calibration by assuming a rational
  functional form (Equation \ref{eqn:recal}) and minimizing the
  deviations with respect to O3N2 metallicities. The resulting new N2
  metallicities show no systematic offsets (right panel) and we use
  them for the remainder of the paper.  \label{fig:o3n2_n2}}
\end{figure*}

\section{Main results}

%
\subsection{Mass--metallicity relations and the systematic differences
  in local metallicity calibrations}

To set the stage for subsequent analysis, we start by inspecting the
\mz\ plots (Figure \ref{fig:mzr1}), which compare the locus of SDSS
galaxies with those from KBSS and MOSDEF samples. Left and middle
panels of Figure \ref{fig:mzr1} are based on N2 and O3N2 line ratios
converted into metallicity using the calibrations of \citet{pp04},
which were derived using mostly direct-method abundances of HII
regions in local galaxies.  For N2, \citet{pp04} provide linear and
cubic calibrations. Figure \ref{fig:mzr1}, (left panel) uses linear
calibration following KBSS and MOSDEF
studies\citep{steidel14,sanders15}.

The first point to be emphasized from Figure \ref{fig:mzr1} is that
the \mz\ relations based on N2 and O3N2 metallicity indicators are not
consistent even for SDSS galaxies, even though \citet{pp04}
calibrations have been based on local galaxies.  Most notably, N2
metallicities at higher mass ($\log M_*\gtrsim10$) are 0.11 dex lower
than O3N2 metallicities. There are several reasons for the local
mismatch, which can be best appreciated from Figure \ref{fig:o3n2_n2},
where the metallicities of SDSS galaxies based on O3N2 and N2
indicators are compared directly. First, the {\it linear} N2
calibration is too crude to capture the saturation of N2 at high
metallicities, and thus leads to a diverging offset (Figure
\ref{fig:o3n2_n2}, left).  This point is relevant for
12+log(O/H)$>8.6$ and thus mostly affects the local samples which
contain galaxies with such super-solar metallicities.  The
relationship between N2 and metallicity in this regime is somewhat
better described with the cubic N2 calibration of \citet{pp04}, but
there are still significant systematics at high metallicities (Figure
\ref{fig:o3n2_n2}, middle; also \citealt{kewley08}, their Figure
2). Second, \citet{pp04} calibrator sample is relatively small
(10$^2$), so some differences due to the limited accuracy of the
functional fitting are to be expected when the comparison is performed
within a much larger (10$^5$) SDSS sample. Furthermore, the
\citet{pp04} calibrator sample may not be entirely representative of a
typical low-redshift population.  Consequently, even for lower
metallicities (12+log(O/H)$\sim8.4$), the metallicity based on linear
calibration of N2 is somewhat offset with respect to O3N2---it is
$\sim 0.03$ dex higher (Figure \ref{fig:o3n2_n2}, left). An offset of
a similar magnitude, but in the opposite direction, is present for the
cubic N2 calibration (Figure \ref{fig:o3n2_n2}, middle). This
discrepancy at lower metallicities will be important for high-redshift
galaxies where measured metallicities tend to be lower. Thus, the 0.13
dex systematic difference between N2 and O3N2 metallicities of
$z\sim2.3$ samples \citep{newman14,zahid14b,steidel14}, which is
usually attributed to evolution in one or both calibrations, is
actually in part ($\sim 1/4$) due to the mismatch in local
calibrations.

In order to be able to separate possible systematics arising from
evolution (i.e., the inapplicability of local calibrations at high
redshift) from those that stem merely from mismatch of the local
metallicity scales, we recalibrate the N2 relation to match the
metallicities resulting from \citet{pp04} O3N2 calibration. We take
O3N2 metallicities as fiducial because the O3N2 calibration is not
subject to saturation, so it is more likely to be close to the linear
form assumed in \citet{pp04}. The new N2 calibration is represented
with a rational function that tends to a linear relation for low
values of N2 (low metallicities), and has asymptotic behavior for high
values. Its analytical form is:

\begin{equation}
12+\log ({\rm O/H})_{\rm N2} = a+b\, {\rm N2} +c/({\rm N2}-d)
\end{equation}
\noindent and ${\rm max}(12+\log ({\rm O/H})_{\rm N2})=e$, where $a,
b, c, d$ and $e$ are the free parameters obtained from the
minimization of the sum of the binned median deviations of orthogonal
offsets of points in Figure \ref{fig:o3n2_n2} from the diagonal (the
1:1 relation). Minimization is performed on binned values in order to
give uniform weight at a range of metallicities. Parameter $d$ represents the
position of the asymptote in N2, while $e$ is the value above which
metallicities are assigned the value $e$. Parameter $e$ signifies the point
above which N2 cannot provide reliable information on metallicity
(except that it is high). The relation with the best-fitting
parameters is:

\begin{equation}
12+\log ({\rm O/H})_{\rm N2} = 8.50+0.37 {\rm N2} -0.15/({\rm N2}+0.10)
\label{eqn:recal}
\end{equation}
\noindent and ${\rm max}(12+\log ({\rm O/H})_{\rm N2})=8.86$. For
comparison, \citet{pp04} linear calibration has parameters $a=8.90$,
$b=0.57$, $c=d=0$. The comparison of new N2 metallicities with O3N2 is
shown in Figure \ref{fig:o3n2_n2} (right). Systematic offsets are
removed. In the remainder of the paper we will only use recalibrated
N2 metallicities.

We return to the \mz\ relation, but now show the plot based on
recalibrated N2 (Figure \ref{fig:mzr1}, right panel). SDSS loci of N2
and O3N2 metallicities now agree better. As expected, there is less
change in trends of $z\sim2$ samples, except that the total change in N2
metallicity from low-mass end to high-mass end is now 
greater, i.e., the \mz\ trend is steeper.

\begin{figure*}
\epsscale{1.1} \plotone{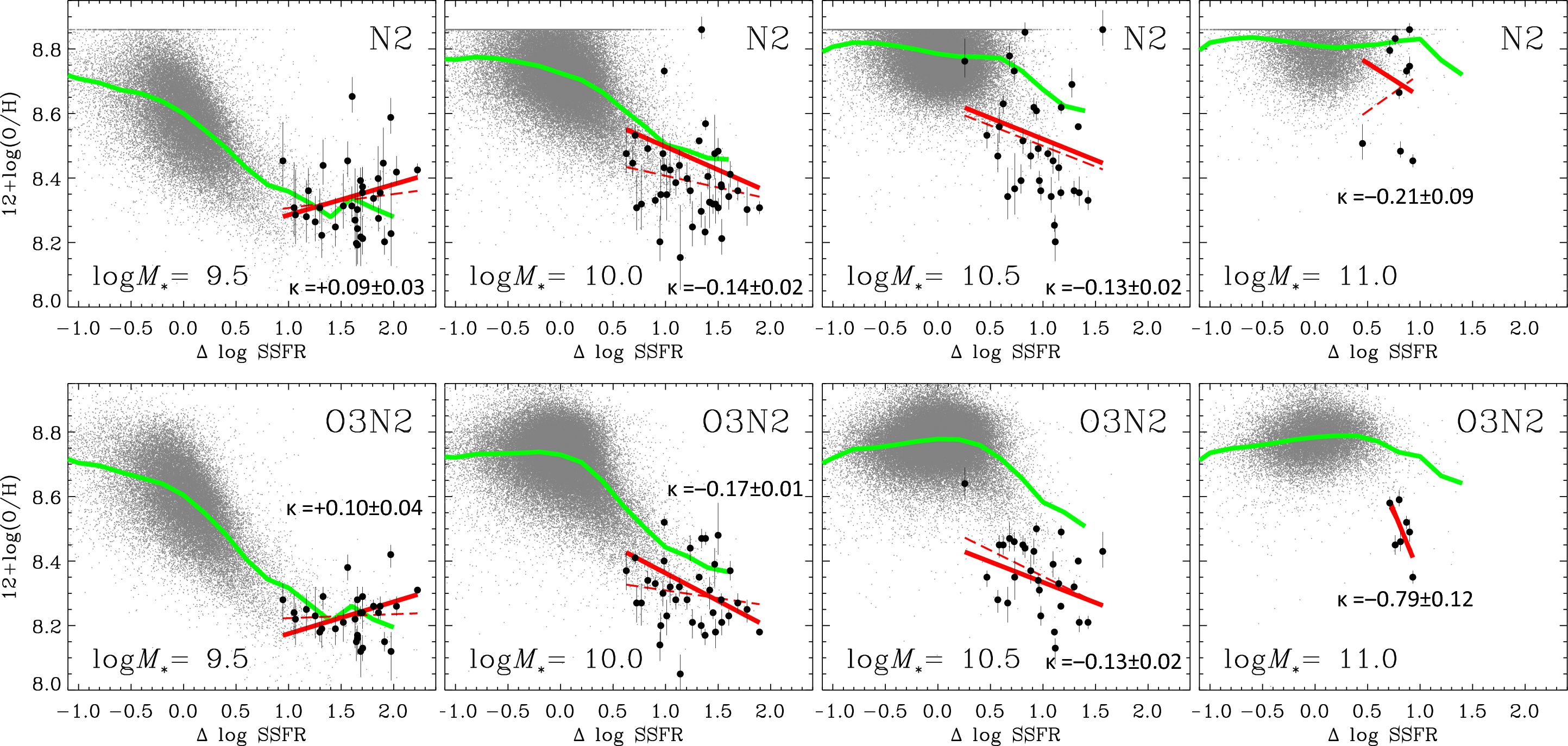}
\caption{\citet{s14} non-parametric analysis technique for studying
  the \mzs\ relation.  Metallicities of local and high-redshift
  galaxies are shown as a function of the relative SSFR (SSFR offset
  from the local star-forming sequence), in four 0.5 dex-wide mass
  bins. Metallicities are derived using two methods: O3N2
  metallicities (lower panels) use \citet{pp04} calibration, while N2
  metallicities (upper panels) are based on our recalibration that
  forces agreement with O3N2 for local galaxies (Figure
  \ref{fig:o3n2_n2}). Recalibration also caps N2 metallicities when N2
  saturates. Black dots represent $z\sim 2.3$ galaxies from KBSS
  \citep{steidel14}, in comparison with SDSS galaxies (grey
  points). High-redshift data show a statistically significant
  anti-correlation at $\log M_*=10.0$ and 10.5 (panels B and C), as
  indicated by the red line (linear weighted fit). Formal slopes and
  their errors are indicated, but whether they are actually
  statistically significant given the sample size requires additional
  consideration (Section 4.2). Visually the correlations appear less
  robust because the points with larger metallicity errors contribute
  to the scatter. Dashed lines show unweighted linear fits. Green
  lines are binned averages for SDSS galaxies (0.2 dex wide bins). For
  both indicators, SDSS and KBSS trends agree in the lowest mass bin,
  but become offset in subsequent bins, with O3N2 offset being
  somewhat larger than N2 offset. \label{fig:dssfr_z1}}
\end{figure*}

We now focus on high-redshift \mz\ trends in comparison with the
local. The most striking difference is that the average N2 line ratios
of the most massive galaxies ($\log M_*>11.0$) are similar for local
and high redshift galaxies (albeit there is a large scatter), implying
little metallicity evolution at the highest end, while the local and
$z\sim2$ \mz\ relations are quite offset in the case of O3N2,
suggesting a strong evolution ($\sim 0.25$ dex).  In Sections 4.4 and
5.3 we will discuss possible causes of this relative discrepancy.

Finally, from Figure \ref{fig:mzr1}, we also see that, regardless of
the indicator, the metallicities of $z\sim2$ galaxies do not saturate
at high masses as they do locally, as pointed out in
\citet{steidel14}. KBSS and MOSDEF \mz\ relations generally agree for
a given indicator, but with some differences in details, which will be
discussed in Section 5.2. The average trends of KBSS and MOSDEF
samples are based on galaxies that are individually detected in
requisite lines (Section 2). For MOSDEF, we do not show the average
\mz\ trends below $\log M_*\sim10$, where many individual galaxies are
not detected in [NII]6584 line \citep{kriek14,sanders15}. Otherwise,
KBSS and MOSDEF \mz\ relations based on individual galaxies agree with
the relations based on stacked spectra \citep{steidel14,sanders15}.

\subsection{Specific SFR as a second parameter of $z\sim2$
  mass--metallicity relation}

\mzs\ relation may be present at $z\sim2$ even if it does not follow
the local \mzs\ relation (i.e., is not redshift invariant), and we
therefore first focus on establishing whether (S)SFR is a second
parameter in KBSS sample. Figure \ref{fig:dssfr_z1} applies the
analysis framework for investigating the \mzs\ relation, as introduced
in \citet{s14}. N2 and O3N2 metallicities are examined against the
relative SSFR in four 0.5 dex wide mass bins, for both the SDSS and
KBSS samples. SSFRs for both sampels are relative to typical local
values at that mass. \footnote{We exclude from analysis Q2343-BX231,
  which, with calculated SFR = 500 $M_{\odot} {\rm yr}^{-1}$ is an
  outlier in $z\sim2.3$ SSFR$-M_*$relation, lying 1.4 dex above it,
  and is also an outlier in the $Z$--SSFR relation.} O3N2 is based on
\citealt{pp04} calibration, while N2 comes from our recalibration
(Equation \ref{eqn:recal}).

To test for SFR dependence we perform weighted and unweighted
least square fits to KBSS points in each mass bin (full and dashed red
lines in Figure \ref{fig:dssfr_z1}). Weighted fits produce
statistically significant non-zero slopes $\kappa$ ($Z$--SSFR
correlations) in all mass bins for both metallicity indicators, except
for $\log M_*=11.0$ bin for N2 ($\kappa=-0.21\pm0.09$). The values of
slopes are given in Figure \ref{fig:dssfr_z1}. In general, the slopes
of KBSS galaxies are comparable to those of SDSS galaxies at the same
mass and SSFR. More specifically, like SDSS galaxies, KBSS sample
shows an anti-correlation between $Z$ and SSFR in $\log M_*=10.0$ and
10.5 bins ($\kappa=-0.14\pm0.02$ and $-0.13\pm0.02$ for N2 and
$-0.17\pm0.01$ and $-0.13\pm0.02$ for O3N2), for the two bins
respectively. 

\begin{figure*}
\epsscale{1.1} \plotone{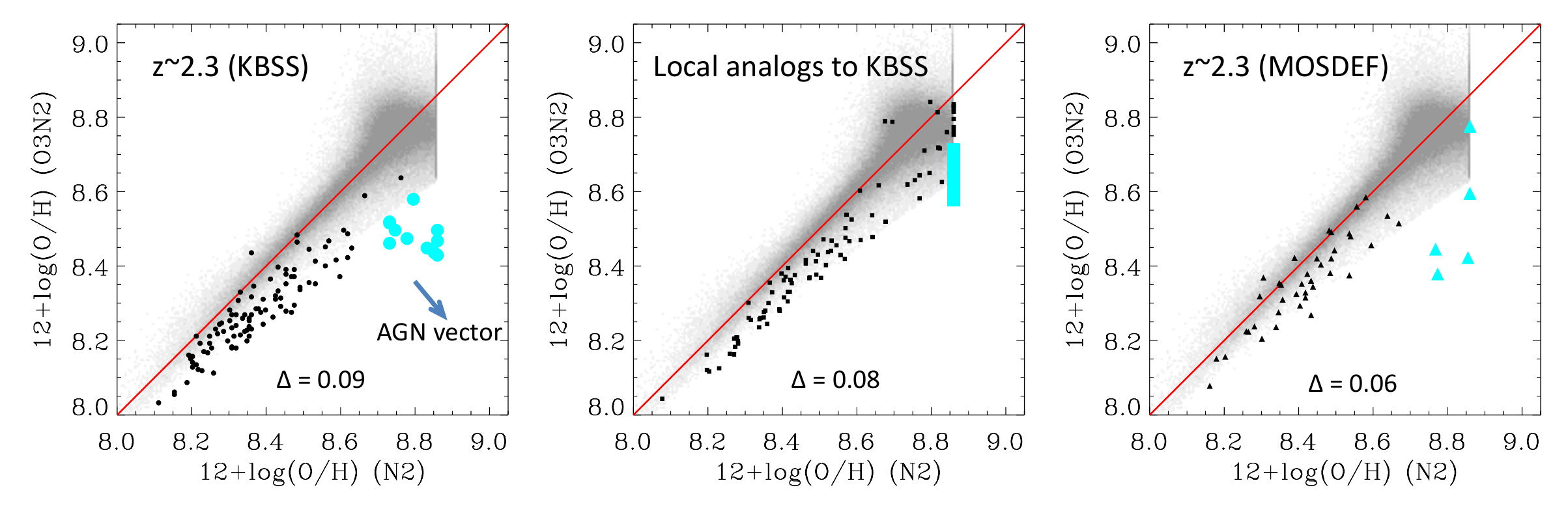}
\caption{Comparison of O3N2 and N2 metallicities of high-redshift
  samples (KBSS, left; MOSDEF, right) and local ``analogs'' to KBSS
  (middle). We define an analog to be a SDSS galaxy with the same SSFR
  and stellar mass as a KBSS galaxy. N2 metallicity is based on our
  recalibration which forces agreement with \citet{pp04} O3N2
  metallicities (Figure \ref{fig:o3n2_n2}) for typical local galaxies
  (grey points). Local analogs to KBSS sample have a similar offset as
  KBSS and MOSDEF samples, validating the \mzs\ analysis approach that
  is based on {\it relative} comparison of metallicities of
  high-redshift and local galaxies of the same mass and SSFR. Possible
  AGN (N2$>-0.5$ and lying above the \citealt{k03c} line) are shown as
  cyan symbols and were excluded from calculating the average offset
  in N2 metallicity ($\Delta$), shown in each panel. The arrow in the
  left panel shows the direction and a possible magnitude of AGN
  contamination vector affecting the cyan
  points.\label{fig:o3n2_n2_highz}}
\end{figure*}

The formal slope errors do not take into account the uncertainties
arising from small sample size. Therefore, to assess the statistical
significance of anti-correlations we perform two additional tests: (a)
we refit 100,000 bootstrapped samples and (b) we perturb the
measurements by randomly drawing from a gaussian with $\sigma$ equal
to the reported metallicity error and by 0.15 dex for SSFR, and refit
the trends 100,000 times. In both tests we perform weighted fits. At
$\log M_*=10.0$, the bootstrap test yields an anti-correlation in 94\%
of cases for N2 (99\% for O3N2), while perturbed fits have a negative
correlation in 99\% of cases (100\% for O3N2). Results are similar for
$\log M_*=10.5$ bin: anti-correlation is present in 90\% (96\% for
O3N2) of bootstraps and 91\% (93\% for O3N2) of perturbed fits. In the
highest mass bin, the trend of KBSS galaxies is not statistically
significant (there is a similar number of correlated and
anti-correlated bootstraps). Currently the sample size in this bin is
too small to establish if the metallicities of $z\sim2$ galaxies, like
the local galaxies of similar mass, lack the dependence on SSFR. In the
lowest mass bin KBSS galaxies do not show an anti-correlation with
SSFR, but neither do SDSS galaxies with such masses and SSFRs. While
the overall SSFR dependence in SDSS at $\log M_*=9.5$ is quite strong,
it appears to reach a low-metallicity plateau for SSFRs 1 dex or more
above the main sequence. KBSS galaxies, which have such high SSFRs,
also appear to have nearly contant metallicities.

We note that the visual impression of the significance of some of the
correlations differs from the results of the formal analysis that,
unlike by-eye estimate, takes metallicity errors into account. Indeed,
the significance of slopes having a non-zero value when fitting is
performed without weighting by metallicity error (dashed lines in
Figure \ref{fig:dssfr_z1}) is low, except for O3N2 galaxies in $\log
M_*=10.5$ bin ($\kappa=-0.16\pm0.06$). For the same reason (ignorance
of weights), the visual impression of the vertical position of the
best-fit linear trend does not always agree with the its position
based on the weighted fit (e.g., $\log M_*=10.0$ bin for N2).

Overall, we conclude that SSFR does appear to be a second parameter in
$z\sim2$ \mz\ relation, at least in the range of masses where such
dependence is clearly present in the local relation.

\subsection{Redshift invariance of the mass--metallicity-SFR relation: standard assumptions}

We now turn our focus to the question of the invariance of \mzs\
relation. In this section we will approach the analysis with standard
assumptions regarding the local comparison sample and the
interpretation of high-redshift line ratios. In subsequent sections we
will investigate the effects of possible AGN contamination of
high-redshift line ratios, and of high-redshift selection effects due
to high [OIII] sensitivity threshold. 

If the \mzs\ relation is redshift invariant, then at any given mass
and SSFR, the average metallicities of SDSS galaxies should be
statistically consistent with those from KBSS. Therefore, Figure
\ref{fig:dssfr_z1} displays N2 and O3N2 line ratios as
metallicities. However, significant discrepancies between N2 and O3N2
metallicities of high-redshift galaxies were found when {\it local}
calibrations were used to derive them
\citep{newman14,zahid14b,steidel14,sanders15}, suggesting ``evolution'' in one
or both indicators. These offsets can be seen in Figure
\ref{fig:o3n2_n2_highz} for KBSS (left) and MOSDEF (right) samples. It
may thus appear that possible ``evolution'' of metallicity
calibrations would preclude the test of redshift invariance of an
\mzs\ relation. However, Figure \ref{fig:o3n2_n2_highz} (middle panel)
shows that, remarkably, similar offset between N2 and O3N2-inferred
metallicities is also present in {\it local} high-SSFR galaxies (we
specifically show what we call the local ``analogs'' of the KBSS
sample, see Section 3). We interpret this to mean that {\it the
  critical limitation for the application of local calibrations at
  high redshift is not so much that they are local, but rather that
  they are based on {\rm typical} local galaxies (or HII regions in
  such galaxies)}.  Therefore, the {\it relative} comparison of line
ratios of high-redshift galaxies and local galaxies of similar SSFR,
which is at the essence of establishing the invariance of any
potential \mzs\ relation, can be carried out regardless of the
uncertainties involving the conversion of line ratios into absolute
metallicities, i.e., without having to decide if it is N2 or O3N2 (or
both) indicator for which the ``local'' calibration is off at high
redshift ({\it and} for local galaxies with high relative SSFRs). With
this important conclusion in hand, we proceed with the analysis.

Figure \ref{fig:dssfr_z1} compares metallicities of SDSS (grey dots
with binned averages shown with green line) and KBSS galaxies (black
points with linear fits shown by red line), using N2 (upper panels)
and O3N2 (lower panels) metallicity indicators. We notice that SDSS
galaxies extend into the range occupied by KBSS, i.e., in each bin
there do exist local galaxies with SSFRs as high as those in
$z\sim2.3$ sample. While such galaxies are extremely rare today, they
are present in the large volume probed by SDSS. Hence, {\it direct
  comparison of local and $z\sim2$ populations is possible, and
  parameterizations of the local \mzs\ relation do not need to be
  extrapolated into the range of SSFRs occupied by high-redshift
  galaxies in order to test for possible evolution.}

Examination of Figure \ref{fig:dssfr_z1} shows that the average trends
of both N2 and O3N2 line ratios for SDSS and KBSS generally agree at
lower masses. However, an offset appears in $\log M_*=10.0$ bin, in
the sense that KBSS metallicities are lower, especially for O3N2. At
$\log M_*=10.5$ the offset grows to $\sim 0.25$ in O3N2, but is
somewhat smaller for N2, where it is compensated by scatter
towards higher metallicities. The offset stays large for O3N2 at $\log
M_*=11.0$, but is again somewhat smaller for N2. Interestingly, even
in bins in which the offsets between average trends are large, there
typically do exist SDSS galaxies with such low metallicities as
KBSS galaxies. 

\begin{figure}
\epsscale{1.1} \plotone{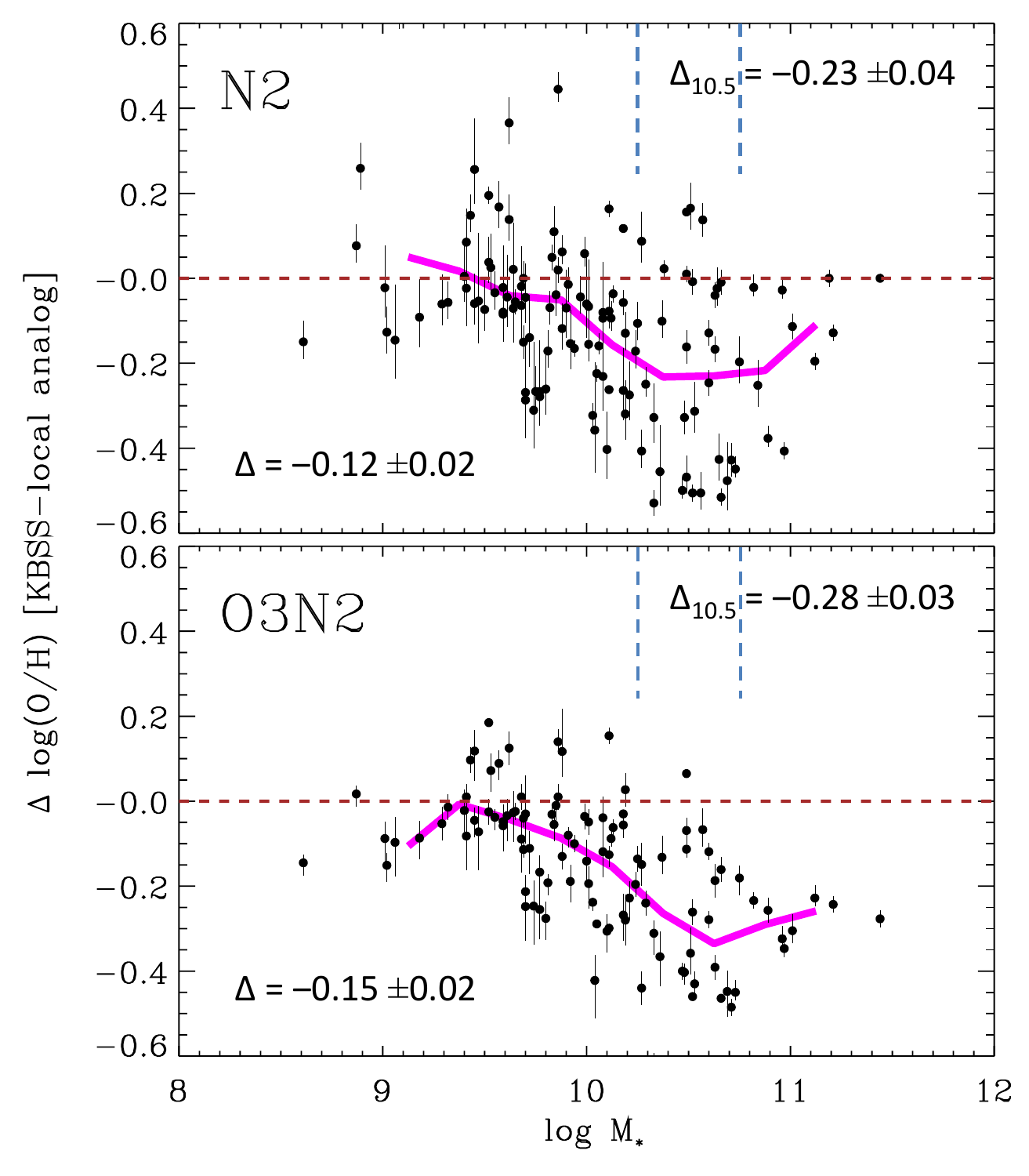}
\caption{Metallicity difference of KBSS galaxies and their local
  ``analogs'' in SDSS using N2 (upper panel) and O3N2 (lower panel)
  indicators. Analog is a galaxy with matching SSFR and stellar
  mass. The offset is mass dependent. The vertical error bar includes
  only the uncertainty of KBSS metallicity, as listed in
  \citet{steidel14}, so it is a lower limit of the full uncertainty of
  the metallicity difference. Colored line show binned averages. Dashed lines
  show the interval centered on $\log M_*=10.5$ used to calculate the
  average offset at high mass
  ($\Delta_{10.5}$).\label{fig:dmet_bptsf}}
\end{figure}

The exact degree of offsets between the SDSS and KBSS trends depends
on how one chooses to treat the data. If metallicity weights are
ignored, as in fitting by-eye, the trend for, e.g., N2 in $\log
M_*=10.0$ bin becomes more offset from the SDSS (dashed line). Thus we
also perform an additional, more direct test of invariance. Instead of
comparing the trends in various mass bins, we now study the difference
in metallicity between KBSS galaxies and their local ``analogs'' from
SDSS (i.e., the galaxy with matching SSFR and $M_*$, see Section 3,
and lying below the \citealt{k03c} AGN demarcation line). We show the
differences in metallicities for each KBSS--analog pair against its
stellar mass (Figure \ref{fig:dmet_bptsf}). Redshift-invariant
relation requires a lack of systematic offsets. We see that this is
the case below $\log M_*\sim 9.7$, but above it the systematic
difference appears, confirming the results from $Z$--SSFR analysis. In
a 0.5 dex interval centered at $\log M_*=10.5$ the average difference
is $-0.23\pm0.04$ dex for N2 and $-0.28\pm0.03$ dex for O3N2 (errors
of offsets were obtained from bootstrapping resampling; they are not
standard deviations).

\begin{figure}
\epsscale{1.1} \plotone{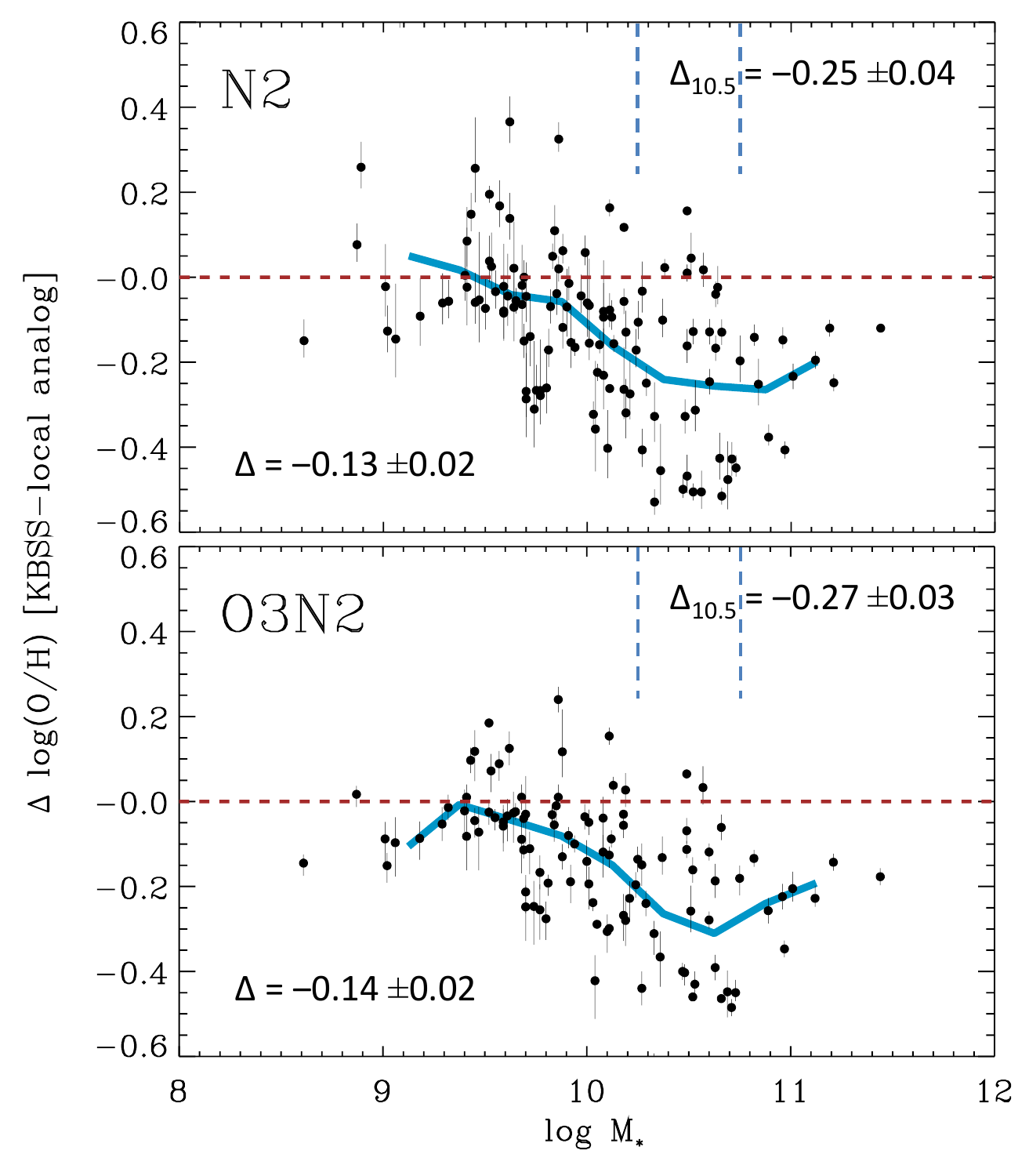}
\caption{Metallicity difference of KBSS galaxies and their local
  ``analogs'' in SDSS. Same as Figure \ref{fig:dmet_bptsf}, except
  that we now apply a correction to galaxies whose line measurements
  may be affected by an AGN contribution (cyan dots in Figure
  \ref{fig:o3n2_n2_highz}, left). The overall effect of the correction
  is relatively subtle, but it does produce more consistent offsets
  for N2 and O3N2 indicators, for the entire sample ($\Delta$), and at
  higher mass ($\Delta_{10.5}$). \label{fig:dmet_agncorr}}
\end{figure}

\begin{figure*}
\epsscale{1.15} \plotone{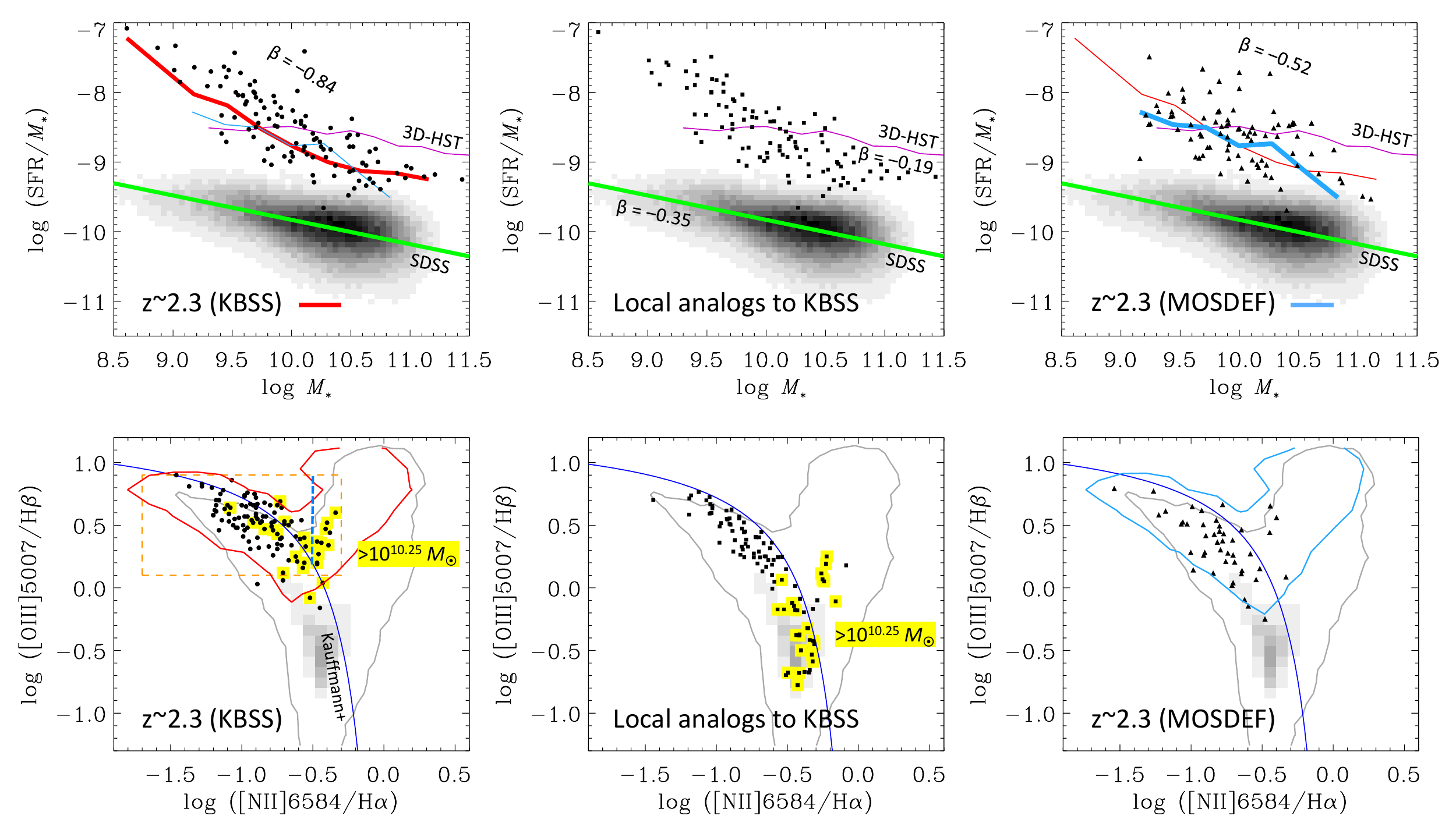}
\caption{Comparison of $z\sim2.3$ KBSS (left), their local SDSS
  analogs (middle) and $z\sim 2.3$ MOSDEF samples (right), in terms of
  SSFR--$M_*$ plane (upper row), and the BPT diagrams (lower
  row). Green lines in the upper row represent a linear fit to the
  SDSS SF sequence (Equation \ref{eqn:ssfr_mass},
  greyscale). High-redshift samples, especially KBSS, have steep
  slopes of SSFR vs.\ mass compared to local galaxies, or even
  compared to other surveys at that redshift (3D-HST,
  \citealt{whitaker14}). High-redshift surveys cannot detect emission
  lines of potential galaxies in the lower part of the BPT diagram
  (red and blue contours show SDSS galaxies with [OIII] and $\ha$
  luminosities above KBSS and MOSDEF detection limits). The lower part
  of the BPT diagram is where locally the more massive galaxies are
  found (yellow dots in left and middle lower panel have $\log
  M_*>10.25$). Dashed vertical line in lower left panel shows
  separates on the right possible AGN in KBSS. Rectangle in the lower
  left panel displays the selection of SDSS galaxies that mimics the
  region detected and occupied by KBSS sample, which admits some
  galaxies that lie above the local AGN demarcation line (blue curve,
  \citealt{k03c}). \label{fig:bpt}}
\end{figure*}

\subsection{Redshift invariance: AGN contamination}

From the analysis presented so far we would conclude that the \mzs\
relation is altogether not invariant, because of the systematic
differences in metallicities at larger masses. The differences tend to be
larger using the O3N2 than N2 indicator. This discrepancy could be the result of
unrecognized AGN contribution to emission lines. Namely, AGN
contribution moves a galaxy along the AGN mixing sequence in the BPT
diagram \citet{bpt} (i.e., O3 vs.\ N2 diagram), increasing both N2 and
O3. The increase in O3 is larger than in N2, so net O3N2 ratio
increases as well. Larger N2 gives {\it higher} apparent
metallicity, while larger O3N2 gives {\it lower} apparent
metallicity, giving rise to a discrepancy.

In Figure \ref{fig:bpt} (lower left) we show the BPT diagram for KBSS
sample. Some galaxies, especially with larger N2 values (N2$>-0.5$)
are found away from the locus of the rest of the galaxies, along what
may be an AGN spur. These galaxies tend to be more massive. These
possible AGN-contaminated galaxies are shown as cyan dots in Figure
\ref{fig:o3n2_n2_highz} (left), in which we directly compared derived
O3N2 and N2 metallicities. These candidate AGN indeed lie even further
from the 1:1 diagonal than the rest of the KBSS sample. Similar effect
was reported in \citet{newman14}. Moving a galaxy up the AGN mixing
sequence increases O3 and N2 such that $\Delta {\rm O3} \approx 5\Delta{\rm
  N2}$. For $\Delta$O3 = 0.4 dex, which appears as a reasonable value
from inspecting Figure \ref{fig:bpt}, the resulting O3N2
``metallicity'' decreases by 0.10 dex, while the N2 ``metallicity''
increases by 0.12 dex, i.e., galaxy is shifted almost perpendicular to
the diagonal in Figure \ref{fig:o3n2_n2_highz} (arrow in the left
panel). Similar shift is seen in several MOSDEF galaxies (right
panel).

Is it realistic to have unrecognized AGNs in the KBSS (and MOSDEF)
samples? KBSS excluded high ionization species emission-line galaxies
as being contaminated by AGN. This method is only sensitive to more
energetic AGNs (ionization potentials for \ion{C}{4} and \ion{N}{5}
are 48 and 77 eV, respectively, compared to 15 eV and 35 eV for [NII]
and [OIII]).  A similar method applied to the SDSS would not identify
as AGN many of the galaxies on the AGN branch of the BPT diagram. Line
ratio diagnostics may be the only practical method to recognize weaker
AGN even locally.

How does the possible AGN contamination bear on testing the \mzs\
redshift invariance? In Figure \ref{fig:dmet_agncorr} we again show
the difference between the metallicities of KBSS galaxies and their
local analogs as a function of mass, but we now correct the
metallicities of AGN candidates by $-0.10$ dex for N2 metallicity (13
galaxies) and $+0.12$ dex for O3N2 metallicity (11
galaxies). Systematic offsets remain, especially at high masses, but
they are now nearly identical for N2 and O3N2 metallicities ($-0.13$
dex overall and $-0.26$ dex at $\log M_*=10.5$).

\subsection{Redshift invariance: [OIII] sensitivity threshold}

In this section we explore whether the sensitivity of high-redshift
spectroscopic surveys \citep{juneau14} affects the analysis of the
evolution of the \mzs\ relation.

To illustrate the potential issue with sensitivity, we again examine
the BPT diagrams in Figure \ref{fig:bpt} (KBSS, lower left panel;
MOSDEF, lower right panel). Both $z\sim2.3$ samples are confined to
the upper part of the diagram, which is locally dominated by lower
metallicity galaxies. What is interesting is that the regions of the
BPT diagram to which the current high-redshift spectroscopic surveys
are sensitive to (shown with red and blue contours in KBSS and MOSDEF
BPT diagrams, respectively), largely {\it coincide} with the locus of
individual detections. (The sensitivity contours were obtained by
selecting SDSS galaxies having [OIII]5007 and $\ha$ luminosities above
KBSS/MOSDEF survey limits (Section 2), following \citealt{juneau14}.)
If galaxies at $z\sim2$ existed in the lower portion of the BPT
diagram, current surveys would not include them in their samples,
because they require [OIII]5007 line detection.

We can also use local ``analogs'' to illustrate the nature of this
potential issue. Like KBSS and MOSDEF, the local
``analogs''\footnote{In this figure we allow the ``analogs'' to be
  selected regardless of the position in the BPT diagram, i.e.,
  including from among the galaxies lying above the \citet{k03c} AGN
  demarcation line.} to KBSS galaxies are offset upwards in the BPT
diagram (Figure \ref{fig:bpt} , lower middle panel), but unlike the
$z\sim2$ samples, the local ``analogs'' to the KBSS sample
populate the full extent of the star-forming branch in the BPT
diagram, with the majority of the more massive ``analogs'' having low
O3 (yellow points indicate $\log M_*>10.25$). The difference in the
extent of the locus of the ``analogs'' on the BPT diagram with respect
to $z\sim2$ samples is essentially another way of saying that the
\mzs\ relation shows no offset at lower masses (upper part of the BPT
diagram), but becomes increasingly discrepant at higher masses,
because the locally higher-mass galaxies populate the lower regions of the
BPT diagram, while they are (intrinsically, or perhaps because of
sensitivity limits) found only in the upper parts at high redshift. 

\begin{figure*}
\epsscale{1.1} \plotone{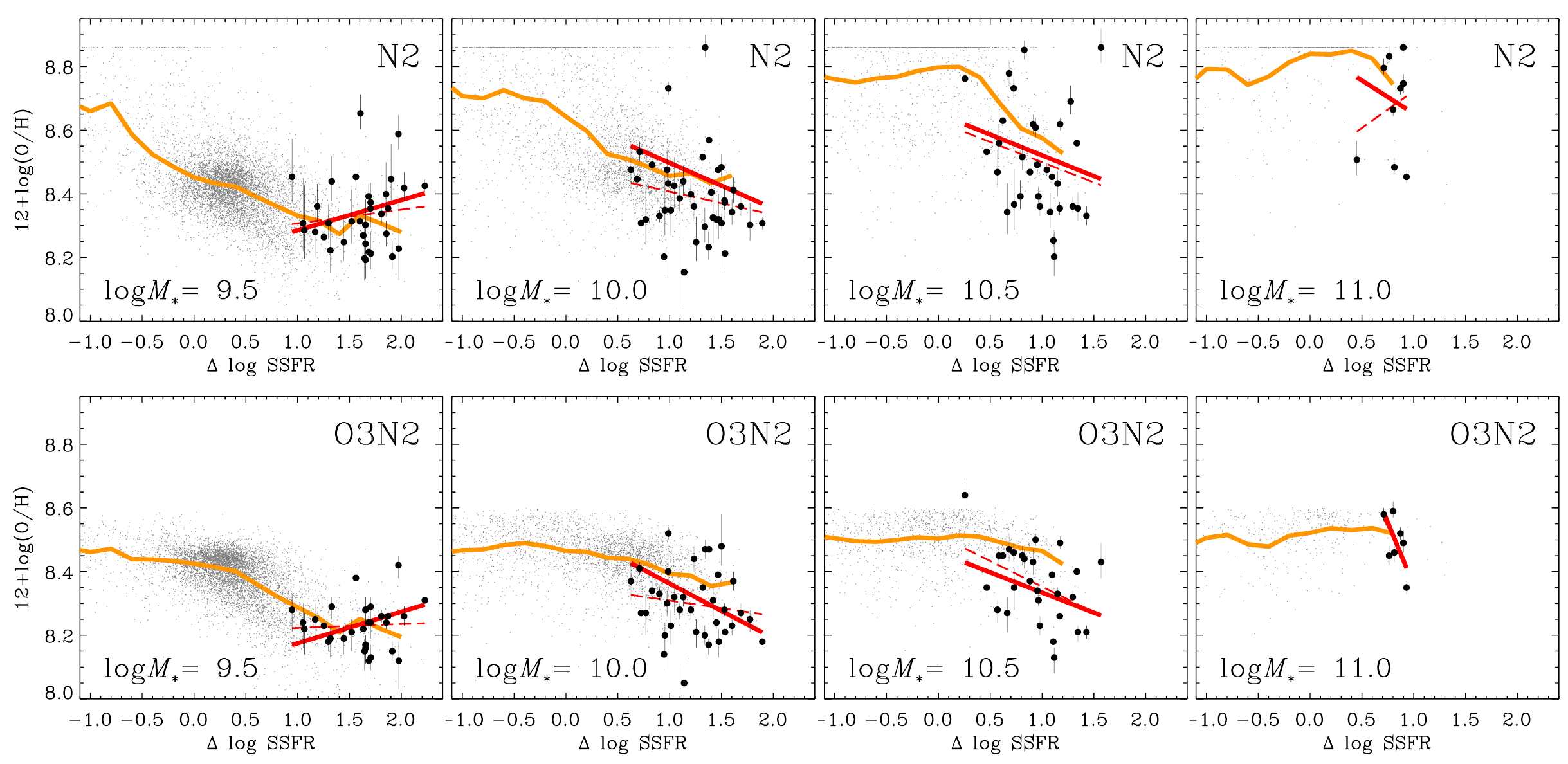}
\caption{Dependence of N2 and O3N2 metallicity on the offset from the
  local star-forming sequence for $z\sim 2.3$ galaxies (black dots)
  and SDSS galaxies (grey points) selected from a similar region of
  the BPT diagram as occupied by high-redshift samples (rectangle in
  Figure \ref{fig:bpt}, lower left). Selection is modified from the
  standard one in order to allow for the possibility that
  high-redshift surveys miss galaxies in regions of the BPT diagram
  that they are not sensitive to, and to allow for the possibility of
  AGN contamination. The agreement between line ratio (metallicity)
  trends is improved, i.e., in this scenario the redshift invariant
  \mzs\ relation would be viable. However, other evidence disfavors
  the existence of a $z\sim2$ high-metallicity population. See also
  Figure \ref{fig:dssfr_z1} caption, except that binned averages for
  SDSS are now shown as orange lines. \label{fig:dssfr_z2}}
\end{figure*}

\begin{figure}
\epsscale{1.1} \plotone{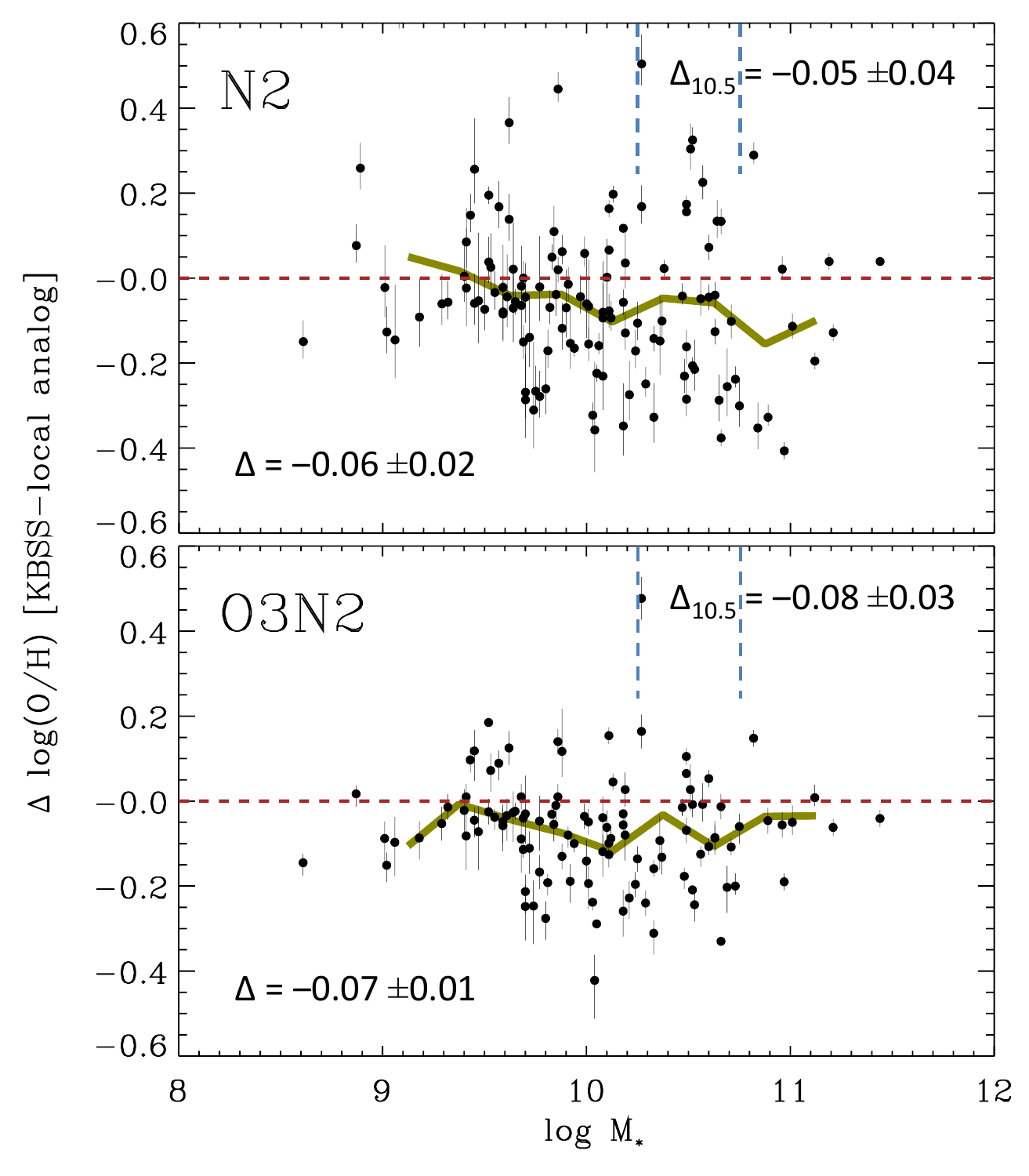}
\caption{Metallicity difference of KBSS galaxies and their local
  ``analogs'' in SDSS. Analogs are now selected from among the SDSS
  galaxies that occupy a similar region of the BPT as high-redshift
  samples, in order to account for the possibility that high-redshift
  surveys miss galaxies in regions of the BPT diagram that they are
  not sensitive to, and to allow for AGN contamination. The offsets
  are reduced compared to standard selection/assumptions (Figure
  \ref{fig:dmet_bptsf}), but are not entirely
  eliminated. \label{fig:dmet_highz}}
\end{figure}

To understand the possible consequences of the insensitivity of
high-redshift surveys to galaxies with low O3 ratios (high
metallicities), we repeat the analysis, but now select SDSS galaxies
only from the region of the BPT diagram accessible to, and occupied by
KBSS galaxies ($0.1<{\mathrm O3}\leqslant0.9$, and $-1.7<{\mathrm
  N2}<-0.3$, shown with the rectangle in Figure \ref{fig:bpt}, lower
left). This selection box will also admit some SDSS galaxies that are
found above the AGN demarcation line, therefore implicitly
``correcting'' for the possibility that $z\sim2$ samples are also
affected by AGN contribution. We therefore do not apply any explicit
AGN correction as we did in Section 4.4.

The new set of $Z$-SSFR plots is presented in Figure
\ref{fig:dssfr_z2}. The offset between SDSS and KBSS trends has
generally reduced compared to Figure \ref{fig:dssfr_z1}. The large
offset in the $\log M_*=11$ bin is eliminated, and is greatly reduced
in the $\log M_*= 10.0$ and 10.5 bins. The improved agreement arises
primarily because the local galaxies with high metallicity, which the
$z\sim2$ surveys would not be sensitive to, are now excluded from the
comparison.  These galaxies pushed the average metallicity trend of
SDSS galaxies upwards.

We also repeat the analysis in which we directly compare the
metallicities of KBSS galaxies with their local ``analogs'', except
that we select the analogs from the rectangle in Figure \ref{fig:bpt},
lower left. The average trends of metallicity difference in Figure
\ref{fig:dmet_highz} confirm that the discrepancies at higher masses
are significantly reduced, although they are not completely
eliminated. At $\log M_*\sim10.5$ the offset is now $-0.05\pm0.04$ for
N2 and $-0.08\pm0.03$ for O3N2. N2 and O3N2 offsets are
comparable. The overall offset (at all masses) is now $-0.06\pm0.02$
with N2, and nearly identical with O3N2.

The above analysis demonstrates that our ability to constrain the
metallicity evolution at higher masses hinges on an implicit
assumption used in all studies so far that there do not exist
significant populations of high-redshift galaxies in regions of the
BPT diagram that are not accessible to observations (low O3
values). May such high-redshift populations exist? The fact that the
ISM conditions at $z\sim2$ appear to be somewhat more extreme than
they are in typical local galaxies \citep{coil15} does not preclude
populating the lower (high-metallicity) portion of the BPT
diagram. This is evident from the position of local analogs in Figure
\ref{fig:bpt} (middle lower panel), which follow the ``more extreme''
location of KBSS galaxies in the upper part of the BPT diagram, but
nevertheless also populate the lower portions. It is obviously
important to understand if observational threshold actually produce
any such biases. Because we do not have access to original KBSS and
MOSDEF data, we base this discussion on the inferences from published
results.

\citet{shapley15} tested whether non-detections create a bias in the
BPT distribution for the MOSDEF sample. They produced {\it
  average}-stacked spectra, in mass bins, that included both the
detections and the non-detections. The stacked spectra had line ratios
that placed them in the region of BPT values shared by the detections
(O3$>0.2$, their Fig.\ 2, left). Because the line ratios are
distributed log-normally, {\it median} stacking of spectra would be
preferred to average stacking, however, the results of such exercise
produce similar results (A. Shapley, priv.\ comm.) A more stringent
test would involve stacking only the non-detections. However, even the
stacked spectra of non-detections are unlikely to fall in the lower
part of the BPT diagram because the majority of non-detections in both
KBSS and MOSDEF are galaxies with weak [NII], and not with weak
[OIII]. Furthermore, the detection fraction tends to be high ($>80$\%)
at higher masses. Therefore, it appears that high-redshift samples,
modulo some other sample selection bias (Section 5.1), are not biased
by high [OIII] detection thresholds, i.e., they are probably not
missing a significant population of high-metallicity
galaxies. Consequently, the fact that the position of $z\sim2$ samples
in the BPT diagram matches the regions that these surveys are
sensitive to (Figure \ref{fig:bpt}) appears to be a mere coincidence,
and a gap between local and high-redshift \mzs\ relations cannot be
attributed to sensitivity issues. Nevertheless, we suggest that future
studies should consider this issue.

\section{Open questions}

%
\subsection{Are KBSS and MOSDEF spectroscopic samples representative
  of star-forming galaxies at $z\sim2$?}

In order to robustly assess the evolution of chemical enrichment from
$z\sim2$ to the present day, one requires both the high-redshift and
the local samples to represent typical star-forming galaxies at that
redshift. It is currently uncertain whether this is the case for
high-redshift spectroscopic samples. The distribution of KBSS and
MOSDEF data in SSFR--$M_*$ plane is different from what is
expected. The slope of the star-forming sequence of KBSS sample is
very steep ($\beta=-0.84$), significantly steeper than $\beta=-0.35$
slope of the local galaxies (Equation 2; green line in the upper
panels of Figure \ref{fig:dssfr_z2}). This disagrees with most results
that suggest that the slope at $z\sim2$ is similar or shallower than
the local slope \citep{speagle14}. Recently, \citet{whitaker14} have
determined the star-forming sequence at $z\sim2.3$ based on an
extensive dataset from 3D-HST survey (purple line in Figure
\ref{fig:bpt} upper panels).  As expected, the slope of the
3D-HST main sequence is somewhat shallower than the local one (we get
$\beta=-0.19$), in sharp contrast to KBSS. At $\log M_*<9.7$ the
3D-HST sequence is lower than the KBSS sequence (but not MOSDEF),
probably due to the KBSS's UV selection, as mentioned in
\citet{steidel14}. However, the situation reverses at $\log M_*>9.7$,
where both KBSS and MOSDEF have {\it lower} SSFRs than 3D-HST, up to
0.4 dex at $\log M_*\approx11$. This high-mass offset has not been
discussed in the literature so far. Altogether, it appears that the
$z\sim2$ spectroscopic samples target somewhat more intense
star-formers at lower mass and more quiescent galaxies at higher
mass. This potential bias does not directly affect for our \mzs\
relation analysis because we are comparing SDSS and KBSS galaxies at
the same specific SFR. Nevertheless, it is important to fully
understand if spectroscopic samples are representative of underlying
star-forming population at $z\sim2$, especially at high mass, where we
find that the \mzs\ relations have an offset.

\subsection{Are there systematic differences between KBSS and MOSDEF line
  ratios?}

The \mz\ relations produced using KBSS and MOSDEF data on individual
detections agree very well (Figure \ref{fig:mzr1}), except above $\log
M_*=10.2$ for O3N2, where MOSDEF metallicities start to diverge from
KBSS ones, the latter being lower. 

The discrepancy between MOSDEF and KBSS measurements has been
discussed in \citet{shapley15} in terms of the position of the
galaxies in the upper portion of the star-forming branch in the BPT
diagram. Here we perform similar analysis, but quantify the
differences between the samples by finding the value O3$_{\rm med}$,
such that the curve:
\begin{equation}
{\rm O3} = \frac{0.61}{{\rm N2}+0.08}+{\rm O3_{med}}
\end{equation}
\noindent divides the sample in half. This function has only one free
parameter: O3$_{\rm med}$. The values of the other two parameters are
fixed to the values used by \citet{kewley13} to describe the
star-forming locus at $z=0$. \citet{steidel14} have allowed all three
parameters to be free, but we find that the single-parameter form with
other parameters fixed to local values actually better describes the
high-redshift samples after possible AGN-contaminated galaxies are
taken out. Furthermore, a single-parameter expression, i.e., a shift
in the O3 direction, makes it easier to compare different samples. For
local galaxies \citet{kewley13} find O3$_{\rm med}$=1.1. For high
redshift samples and their local analogs we focus only on the upper
left part of the BPT diagram (N2$<-0.5$, O3$>0.0$), thus excluding
possible AGN contamination (see also Figure \ref{fig:o3n2_n2_highz})
and outliers. For KBSS we obtain O3$_{\rm med}$=1.32, i.e., its
star-forming branch lies some 0.2 dex higher in the BPT diagram than
the local one. For MOSDEF, we get O3$_{\rm med}$=1.20, which, while
higher than the local value, is significantly lower than the KBSS
value. Local analogs to KBSS yield O3$_{\rm med}$=1.28, which is
closer to KBSS than MOSDEF, and also much higher than the typical
local galaxies. That the local galaxies with high relative SSFRs were
offset in the BPT diagram was previously found in
\citet{brinchmann08}.

\citet{shapley15} ascribe the difference between the position of KBSS
and MOSDEF samples in the BPT diagram sample selection: UV selection of
KBSS sample as opposed to the optical selection of MOSDEF. However, it
is not clear whether selection is the main culprit. Inspecting the
SSFR--$M_*$ distributions of the two samples (Figure \ref{fig:bpt}
upper left and right), they appear similar, except at $\log M_*<9.5$
where KBSS sample has higher SSFRs and extends to lower masses. Even
when we restrict the KBSS sample to mass range that both KBSS and
MOSDEF appear to cover in similar fashion ($9.5<\log M_*<10.5$), we
still obtain O3$_{\rm med}$=1.29, a significantly higher value than
for MOSDEF \footnote{We cannot perform equivalent, mass-range limited
  determination for MOSDEF because we do not have matched tables of
  line ratios and masses.}. Local analogs in the same mass range yield
O3$_{\rm med}$=1.27, i.e., very close to KBSS value. Future studies
should be able to pin down the source of the discrepancy. The analysis
presented here suggests that the cause may not be solely the
differences in sample selection.

We also wish to point out that while O3$_{\rm med}$ are similar for
KBSS and their local analogs, the former have a larger scatter
(presumably because of larger measurement errors), which is why they
more often cross the local AGN demarcation line than their analogs,
which are found close to the line but do not cross it. Indeed, if we
increase O3 values of analogs by the difference in O3$_{\rm med}$
between the analogs and KBSS (0.046 dex), only 6 out of 71 galaxies
from the upper left of the BPT diagram would actually cross the
\citet{k03c} line. This suggests that $z\sim2$ galaxies without AGN
contribution {\it intrinsically} mostly lie below the \citet{k03c}
line. This conclusion would be even stronger for MOSDEF, whose
O3$_{\rm med}$ is {\it lower} than that of KBSS analogs. We conclude
that had the $z\sim2$ line ratios had the same precision as in SDSS,
the \citet{k03c} AGN classification line would not require a shift
greater than 0.05 dex in the vertical direction. The appearances are
significantly different because of the larger measurement error that
contributes to the scatter and because some of the galaxies that are
currently not considered to be AGN probably are AGN. Allowing for
larger observational errors, but still removing the potential AGN may
be achieved using a simple cut in O3 and N2:

\begin{equation}
{\rm N2}>-0.5 {\rm and} {\rm O3}>0.1,
\end{equation}
\noindent which resonates with the conclusions of \citet{coil15} that
were based on the MOSDEF sample.

\begin{figure}
\epsscale{1.15} \plotone{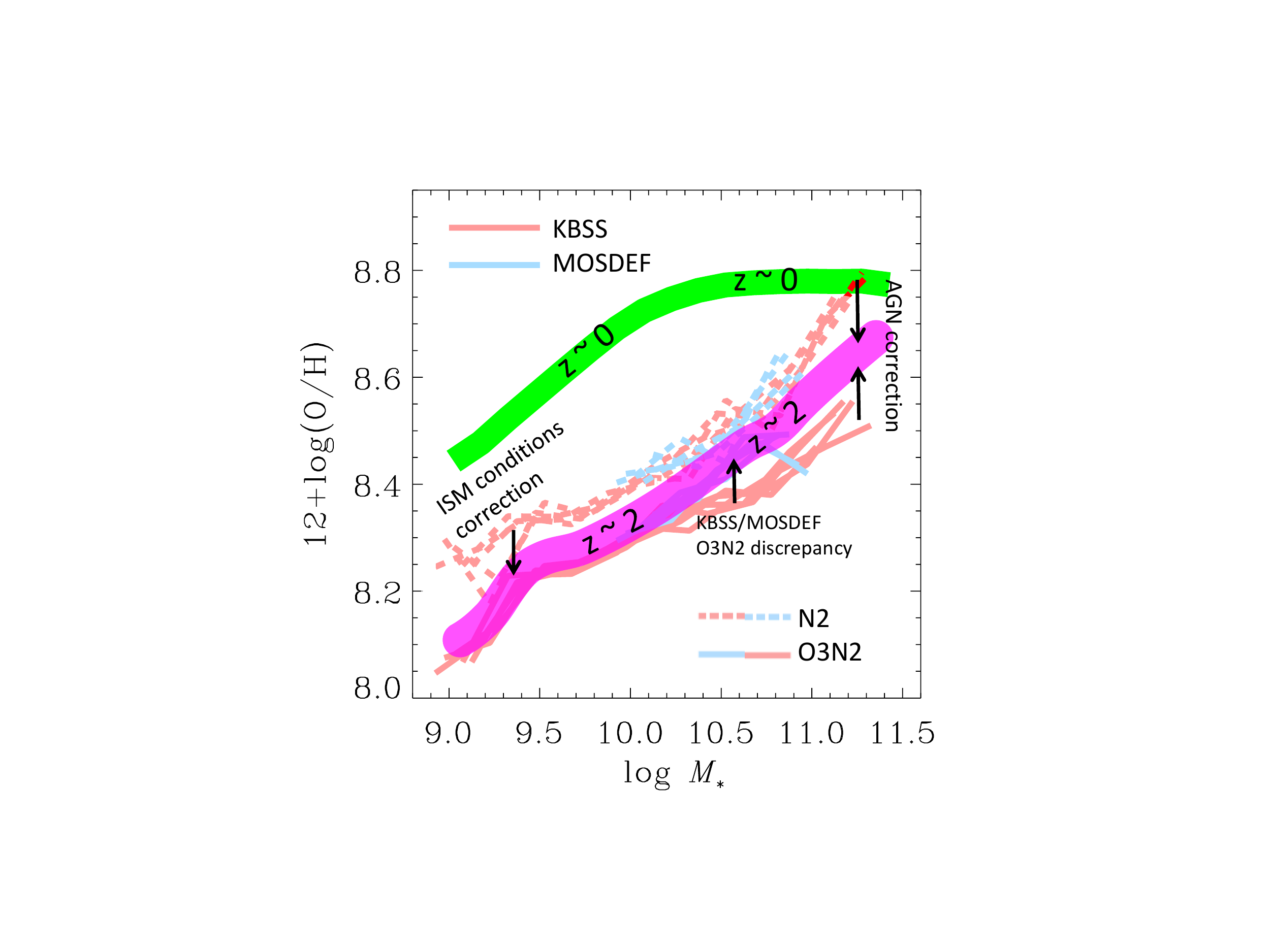}
\caption{Evolution of the mass-metallicity relation from $z\sim2$ to
  today. Concordant high-redshift \mz\ trend (violet band) is a
  qualitative illustration of how to reconcile the \mz\ relations
  based on O3N2 (solid lines) and N2 (dashed lines) metallicity
  indicators, from KBSS (red) and MOSDEF (blue) surveys. To do so we
  consider various systematics discussed in Section 5.3 and labeled in
  the figure. Generally, at lower masses the concordant trend is
  closer to O3N2 because N2 will be more affected by the change in the
  N/O ratio, and at higher masses it is between N2 and O3N2 trends
  because of the possible AGN contamination that boosts N2 but
  suppresses O3N2. \label{fig:mzr2}}
\end{figure}

\subsection{Towards the concordant evolution of mass-metallicity relation}

In this section we will attempt to arrive at a consistent \mz\
relation at $z\sim2$ and discuss its evolution. This requires the
knowledge of absolute metallicities. As discussed in Section 4.3, the
analysis of \mzs\ invariance, being relative, could be carried out even
without the knowledge of absolute calibrations that would be needed to
convert the line ratios of high-redshift galaxies (and their local
analogs) into metallicities. This is not the case if we are interested
in constructing the \mz\ relations. The remaining offset between N2
and O3N2 ``metallicities'' even after we have corrected for a mismatch
that was present in \citet{pp04} relations, and have corrected for
possible AGN contribution (Figure \ref{fig:o3n2_n2_highz}), means that
either or both of the local calibrations does not hold for ISM
conditions in $z\sim2$ galaxies {\it and} their local analogs.

Whether N2 or O3N2 is more affected by the changes in the ISM
conditions is currently a matter of debate. \citet{masters14} and
\citet{shapley15} argue for O3N2 being less affected because $z\sim2$
galaxies appear to follow local galaxies in O3 vs.\ S2 (=
log([SII]6573/$\ha$)) and O32 vs.\ R23 diagrams.  Such diagnostics may
not be particularly relevant in the context of O3N2 vs.\ N2 debate
because using O3, and especially O32, results in comparing
high-redshift galaxies with local galaxies having high ionization
parameters, which, as we have shown, also exhibit the O3N2 vs.\ N2
discrepancy. \citet{steidel14} consider O3N2 to be less biased than N2
based on a very small number of direct metallicities of the local
``green pea'' galaxies, while \citet{liu08} reach a similar conclusion
based on direct metallicities of stacked SDSS galaxies with high O3
values.  On the other hand, \citet{zahid14b} consider N2 to be more
reliable based on the greater sensitivity of O3N2 to changes in the
ionization parameter, which many studies
\citep{brinchmann08,nakajima14,kewley13} assume to be the principal
difference between ISM conditions of local and high-redshift
galaxies. In contrast, \citealt{steidel14} show, using photoionization
models, that the ionization parameter cannot be the only factor
driving the difference. The greater effect of the ionization parameter
on O3N2 may be countered by higher N/O ratio at $z\sim2$ compared to
typical local galaxies of the same metallicity, as tentatively
measured by \citet{masters14}, thus making O3N2 more reliable for
high-redshift studies than N2 in the end.

For the purposes of the remaining discussion we will assume that O3N2
provides more robust metallicities than N2 and that the local O3N2
calibration can be applied at $z\sim2$. Figure \ref{fig:o3n2_n2_highz}
then implies that to correct our recalibrated N2 metallicities for
high-redshift/local analog galaxies one would need to subtract $\sim
0.08$ dex.

Based on all of the inferences made so far we present a possible
evolution of the \mz\ relation from $z\sim2$ to today in Figure
\ref{fig:mzr2}. This figure shows KBSS and MOSDEF metallicity trends
based on both N2 (dashed lines) and O3N2 (solid lines) and then
selects a trend that better fits available evidence (violet band).  At
lower masses ($\log M_*\lesssim10.3$) the final trend is closer to
O3N2 than to N2 because O3N2 indicator is likely more robust for
samples with very high SSFRs.  We allow the final trend to sit
slightly above the O3N2 trends to account for $z\sim2.3$ samples
possibly having atypically high SSFR for that redshift (Section
5.1). As we progress towards higher masses, the possible AGN
contamination affects both O3N2 and N2, but in different
directions. Applying a $\Delta$O3=0.4 dex correction to AGN candidates
(cyan dots in Figure \ref{fig:o3n2_n2_highz}) results in N2 and O3N2
trends that are much closer to each other and roughly between the
uncorrected values.

Comparing the final $z\sim2$ trend now with the local \mz\ relation
(green band) we see that the level of metallicity evolution appears to
be mass-dependent above $\log M_*=10.3$, because unlike the local \mz\
relation, the high-redshift one does not saturate (flatten). Note that
the proper comparison requires the local \mz\ relation to be
constructed in an unbiased way. In particular it is important not to
introduce an [OIII]5007 selection, as it would preferentially remove
the highest metallicity galaxies and could therefore modify the
character of the \mz\ relation at higher masses. At higher masses the
final $z\sim2$ trend approaches that of the local galaxies, but the
gap is not fully closed. At masses below the local plateau, the
$z\sim0$ and $z\sim2$ \mz\ relations have a similar slope, as
advocated in ``universal metallicity relation'' model of \citet{zahid14a}.

\section{Discussion}

%
\subsection{Prior work on $z\sim2$ \mzs\ relation}

\citet{steidel14} and \citet{sanders15} did not detect a significant
secondary dependence of the \mz\ relation on SFR in their analysis of
KBSS and MOSDEF data, while our analysis of some of these same data
does find a dependence on SFR. The likely reason why the SFR
dependence was not found in previous $z\sim2$ studies is because the
analysis methods used in those studies did not fully account for the
fact that the \mzs\ relation is intrinsically not very tight, even
locally where measurement errors are smaller, and that the relation is
driven by changes in SFR relative to the typical SFR at a fixed mass,
rather than absolute SFR \citep{s14}. \citet{steidel14} and
\citet{sanders15} split the sample into high and low SFR and then look
for offsets between respective MZRs. This method will not show a SFR
dependence because it selects by absolute SFR, and because the
information from a limited range of SFRs at a given mass is collapsed
into two closely separated bins. \citet{wuyts14} split their sample
into mass-dependent low/high SFR bins (also mass-dependent SFR
quartiles, R.\ Sanders, priv. comm.), which is more similar to our
methodology, but even the mass-dependent binning is apparently too
crude to uncover a relatively weak SFR dependence, especially in
relatively small samples. \citet{maier14} do not bin their data, so
they tentatively detect the dependence on SFR in their sample of 20
galaxies at $z\sim2.3$. The key to being able to tease out the SFR
dependence in relatively noisy high-redshift data is not to bin by SFR
or any other SFR-related quantity, but to directly look at metallicity
as a function of SSFR (or relative SSFR) and do so separately for
galaxies of different stellar masses.

As pointed out, the ``direct" approach used here and in
\citet{sanders15} for testing redshift invariance of the \mzs\
relation has an advantage over previously applied methods because it
does not rely on extrapolations of parameterizations which approximate
the shape of the surface defined by local galaxies, which do not
adequately capture the behavior of outlier populations such as
galaxies with high SSFRs, and can lead to differing predictions for
metallicities at high redshift \citep{maier14}. Here we have shown
that extrapolations of the local trends are not needed -- direct
comparison to local galaxies that occupy the same part of SSFR--$M_*$
space as $z\sim2.3$ galaxies is possible. The same conclusion was
reached in \citet{sanders15}, while some previous studies (e.g.,
\citealt{maier14,steidel14}) considered the need to extrapolate the
parameterizations of local trends to be essential. However, they based
that need on the range of SDSS SFR values reported in
\citet{mannucci10}, who used {\it fiber} SFRs, which, in addition to
being distance-dependent, are on average 0.6 dex lower than the total
SFRs \citep{s14}.  Many studies have in addition adopted Mannucci et
al.\ parametrization of the local trends to test for \mzs\ invariance
(e.g., \citealt{cullen14,wuyts14,zahid14b}). In contrast to these
studies, \citet{sanders15}, like our study, avoided parameterization,
and used the line ratios from the actual SDSS data, binned by mass and
total SFR, to perform a direct comparison with their MOSDEF
sample. They find that the $z\sim2.3$ metallicities are $\sim 0.1 dex$
lower than SDSS metallicities of galaxies with similar mass and SFR,
similar to our overall result (Figure \ref{fig:dmet_bptsf}). We
present a more nuanced picture where the discrepancy is preferentially
present in more massive galaxies.

Our study also sheds new light on the discrepancy between N2 and O3N2
metallicities at $z\sim2$
\citep{newman14,cullen14,zahid14b,steidel14,sanders15}. We find
evidence for three independent causes. A smaller part ($\sim 0.03$
dex) is due to a mismatch in local calibrations (Section 4.1). The
larger part (0.06--0.09 dex) is due to changed ISM conditions with
respect to typical local galaxies. Figure \ref{fig:o3n2_n2_highz}
demonstrates that such different ISM conditions are not exclusive to
high-redshift galaxies, but are common to their local analogs. Lastly,
and preferentially affecting high-mass galaxies, an additional offset
is due to AGN contamination of [NII] and [OIII] lines, which increases
the relative discrepancy in derived ``metallicities'' by additional
$\sim0.2$ dex (Section 4.4).

\subsection{Implications}

Secondary dependence of the \mz\ relation on SFR at {\it any redshift}
appears as a natural outcome in recent models of galaxy evolution,
both analytical (e.g., \citealt{lilly13,dave12}) and numerical (e.g.,
\citealt{dave11}).  It signifies departure from equilibrium
metallicity at a given mass due to the variations in the gas infall
rate, which also modulates star formation. That we should find
evidence for SFR dependence at $z\sim2$ is therefore somewhat
expected. However, our findings show that discussions of SFR
dependence must be more nuanced. The strength of the SFR dependence in
the local universe obviously depends on the mass, and is very weak or
absent at high mass \citep{ellison08,mannucci10,s14}. This may be
because the local massive galaxies ($\log M_*\approx11$), have by now
reached a ``saturation'' metallicity \citep{zahid14a}, so are more
difficult to perturb chemically by gas infall, even though the infall
would increase the SFR. Current samples at $z\sim2$ are too small to
establish if high-redshift, high-mass galaxies behave in a similar
way. They may not, considering that their metallicities appear to be
below the saturation level.  The results we present here reveal that
the situation may also be more complex at lower masses ($\log
M_*\approx9.5$). There, the overall local SFR dependence is very
strong, but then appears to reach a low-metallicity plateau, such that
the further increase in SSFR does not lead to further decrease in
metallicity. High-redshift galaxies with $\log M_*\sim9.5$, which have
similar SSFRs as the local plateau galaxies also show no
anti-correlation between metallicity and SSFR. Future theoretical
studies should attempt to explain these detailed behaviors.

For the \mzs\ relation to be invariant with redshift requires
additional constraints.  In the ``gas regulator'' model of
\citet{lilly13}, an invariant \mzs\ relation emerges only if the gas
consumption timescale and mass loading of wind outflows are constant
in time (see also \citealt{forbes14}). \citet{lilly13} models with
constant SF efficiency also predict that the metallicity evolution
will decrease with mass, which, in the context of invariant \mzs\
relation, is equivalent to having the SSFR dependence of \mzs\
relation decrease with mass. Mass-dependent evolution of the \mz\
relation is also predicted by the ``universal metallicity relation''
of \citet{zahid14a}, which relates metallicity and gas-richness. More
detailed overview of recent theoretical efforts is given in
\citet{s14}. Establishing whether \mzs\ relation is invariant or not
is therefore important to guide our understanding of the complex
interplay between infall, outflows and SF. Again, our results
challenge simple explanations. We find that local and $z\sim2$ \mzs\
relations are consistent with each other at lower masses ($\log
M_*<10$), but then quickly reach a significant offset
($\sim0.25$). This result suggests that the low-mass local
``analogs'', rare galaxies that are found 0.7 dex or more above the
main sequence, may have a similar mode of SF as the high-redshift
galaxies of the same mass, and are similarly ``unevolved''. High mass
galaxies, on the other hand, tend to be metal-rich today, even when
their SFRs are as high as those of high-redshift galaxies, suggesting
that they owe high SFRs to different processes from those that operate
at high redshift.

\section{Conclusions}

We conclude that barring some selection effect, the \mzs\ relation of
$z\sim2$ galaxies is consistent with the local one at lower masses,
but not at higher masses, so it is overall not redshift invariant.  We
summarize the main conclusions of this study are as follows:

\begin{enumerate}

\item Local ($z\sim0$) galaxies with SSFR and $M_*$ typical of
  $z\sim2.3$ spectroscopic samples exist in SDSS, which allows the
  redshift invariance of the \mzs\ relation to be studied directly,
  using the non-parametric method of \citet{s14} or the related
  ``local analog'' method presented in this paper. Local analog method
  consists of identifying a local galaxy in SDSS (or another
  large-volume spectroscopic survey) whose stellar mass and SSFR match
  closely that of a high-redshift galaxy, and evaluating systematic
  differences between the metallicities of local analogs and
  high-redshift sample (Figures \ref{fig:dssfr_z1} and
  \ref{fig:dmet_bptsf}).

\item The \mz\ relation at $z\sim 2.3$ shows a statistically
  significant dependence on SSFR at intermediate masses ($9.7<\log
  M_*<10.7$), the same mass range where such dependence (i.e.,
  $Z$--SSFR anti-correlation) is seen in local (SDSS) galaxies with
  similarly high SSFRs. Above $\log M_*=10.7$ no conclusions can be
  drawn because of the very small number of such galaxies in the
  sample. Anti-correlation is not present for lower masses ($\log
  M_*\sim9.5$) KBSS galaxies, a behavior which appears to also hold
  for SDSS galaxies with such mass and KBSS-like SSFRs. (Figure
  \ref{fig:dssfr_z1}).

\item \mzs\ relation of $z\sim2.3$ KBSS sample shows no offset with
  respect to the local \mzs\ relation at lower masses ($\log
  M_*\lesssim10$), regardless of whether N2 or O3N2 lines are used to
  derive the metallicities (Figure \ref{fig:dssfr_z1}).

\item An offset between high-redshift and local \mzs\ relations does
  emerge at higher masses, reaching, around $\log M_*=10.6$, $\sim
  -0.2$ dex for N2 and $\sim -0.3$ dex for O3N2 metallicity. The sense
  of the offset is that $z\sim2$ metallicities are lower than what is
  expected from the local \mzs\ relation. \mzs\ relation is therefore
  altogether not redshift invariant (Figures \ref{fig:dssfr_z1} and
  \ref{fig:dmet_bptsf}).

\item This high-mass offset becomes consistent for N2 and O3N2
  ($-0.26$ dex) if we correct some high-mass, high-redshift galaxies
  for the effects of unrecognized AGN contribution. AGN contamination
  is implicated because it boosts both N2 and O3, which, when
  interpreted as metallicity, leads to overestimates based on N2 and
  underestimates based on O3N2. AGN indicators other than the line
  ratios will have difficulty recognizing these galaxies as AGN even
  locally (Figures \ref{fig:dssfr_z1}, \ref{fig:bpt}, and
  \ref{fig:o3n2_n2_highz}).

\item Current high-redshift spectroscopic surveys are biased against
  high-metallicity galaxies because they would have [OIII]5007 lines
  that fall below the [OIII] sensitivity thresholds. This selection
  effect could in principle explain the gap in \mzs\ relations at high
  mass and open up the possibility for a redshift-invariant
  relation. However, the majority of current non-detections have weak
  [NII], rather than [OIII], and few non-detections are at high mass,
  making this scenario for removing the \mzs\ offset unlikely (Figures
  \ref{fig:bpt}, \ref{fig:dssfr_z2} and \ref{fig:dmet_highz}).

\item Local ``analogs'' of high-redshift samples (selected only to
  have the same stellar mass and SSFR) are similarly displaced in the
  upper part of the BPT diagram with respect to the bulk of
  low-redshift galaxies as the high-redshift samples, suggesting that
  the lower-metallicity local analogs may have ISM conditions in
  common with high-redshift populations. The consistency between the
  line ratios of the high-redshift and local analogs supports our
  implicit assumption that equal line ratios at a given SSFR (and
  $M_*$) indicate the same metallicities, even if the absolute value
  of this metallicity may not be accurately given by widely-used local
  calibrations. In other words, the inapplicability of local
  calibrations to high-redshift samples is not due to the fact that
  they are local {\it per se}, but rather that they are largely based
  on {\it typical} local galaxies, and therefore do not account for
  the behavior of outlier populations such as the galaxies with very
  high SSFR for their mass (Figures \ref{fig:bpt} and
  \ref{fig:o3n2_n2_highz}).

\item The discrepancy between N2 and O3N2 metallicities reported at
  $z\sim2$ \citep{zahid14b,steidel14,sanders15} has three independent
  causes. A smaller part ($\sim 1/4$) stems from a local mismatch in
  linear calibrations as given by \citet{pp04}. We remove this offset
  by recalibrating the N2 conversion to match \citet{pp04} O3N2
  metallicities of local SDSS galaxies. Larger part is because of the
  changed ISM conditions at $z\sim2$, as suggested in previous
  studies. However, we find that this offset is present to same degree
  in local ``analogs'' (Item 7). Finally, the largest outliers are
  consistent with being additionally offset due to an AGN contribution
  (Item 5) (Figures \ref{fig:o3n2_n2} and \ref{fig:o3n2_n2_highz}) .

\end {enumerate}

Additional conclusions include:

\begin{enumerate}
\setcounter{enumi}{8}
  
\item KBSS O3 line ratios are on average higher than those of MOSDEF galaxies
  above $\log M_*=10.2$, for reasons that may not be fully accounted for by    
  the differences in the sample selection. We cannot tell which values
  should be more typical at $z\sim2$ (Figure \ref{fig:mzr1}).

\item Intrinsically, the AGN demarcation line at $z\sim2$ probably
  lies no more than 0.05 dex higher than the local empirical
  demarcation line \citep{k03c}. Appearance of $z\sim2$ BPT diagrams
  suggests otherwise because of the scatter from larger measurement
  errors and because some of the galaxies that are not considered to
  be AGN probably are (Figure \ref{fig:bpt}).

\item KBSS, and to some extent MOSDEF, have log SSFR versus $\log M_*$
  distributions (``main sequences'') that are steeper (more
  mass-dependent) than that of $z\sim2.3$ galaxies from 3D-HST survey
  \citep{whitaker14}. In particular, at $\log M_*\gtrsim10$, the
  3D-HST ``main'' sequence is on average $\sim 0.4$ dex higher than
  either KBSS or MOSDEF. While this potential bias in spectroscopic
  samples is unlikely to affect our \mzs\ relation analysis, its
  sources need to be investigated (Figure \ref{fig:bpt}).

\end{enumerate}
 
In addition to presenting many new results, we have highlighted
observational issues that need to be fully understood in future work
before more definitive conclusions on the subject of chemical
evolution can be drawn.

\acknowledgments We thank Ryan Sanders and Alice Shapley for
clarifications regarding their work and useful feedback on the
manuscript. We also thank the referee Simon Lilly for constructive
remarks.


\end{document}